\documentclass[aps,twocolumn,amsmath,amssymb,nofootinbib,tighten, superscriptaddress ,floatfix,prb]{revtex4}
\usepackage{color}
\usepackage{graphicx}
\usepackage{dcolumn} 
\usepackage{bm, bbm}
\usepackage{curves}
\usepackage{epic}
\usepackage{wasysym}
\usepackage{epsf, times}
\usepackage{subfigure}
\usepackage{eufrak}
\usepackage{amsthm}

\def\re#1{(\ref{#1})}

\def\bf#1{\bm{#1}}

\def\atan{\mathop{\mathrm{atan}}}
\def\atanh{\mathop{\mathrm{atanh}}}
\def\acos{\mathop{\mathrm{acos}}}
\def\widebar{\overline}

\newcommand{\be}{\begin{equation}}
\newcommand{\ee}{\end{equation}}
\newcommand{\bea}{\begin{eqnarray}}
\newcommand{\eea}{\end{eqnarray}}

\begin{document}

\title{ 
Spatially Anisotropic Heisenberg Kagome Antiferromagnet}

\author{Andreas P.\ Schnyder}
\affiliation{Kavli Institute for Theoretical Physics,
  University of California,
  Santa Barbara,
  CA 93106,
  USA}

\author{Oleg A.\ Starykh}
\affiliation{Department of Physics, University of Utah,
Salt Lake City, Utah 84112, USA}

\author{Leon Balents}
\affiliation{Kavli Institute for Theoretical Physics,
  University of California,
  Santa Barbara,
  CA 93106,
  USA}

\date{\today}

\begin{abstract}
  We study the quasi-one-dimensional limit of the spin-1/2 quantum
  Heisenberg antiferromagnet on the kagome lattice.  The lattice is divided
  into antiferromagnetic spin-chains (exchange $J$) that are weakly
  coupled via intermediate ``dangling'' spins (exchange $J'$).  Using
  one-dimensional bosonization, renormalization group methods, and
  current algebra techniques the ground state is determined in the limit
  $J' \ll J$. We find that the dangling spins and chain spins form a
  spiral with $O(1)$ and $O(J'/J)$ static moments, respectively, atop of
  which the chain spins exhibit a smaller $O[(J'/J)^2]$
  antiferromagnetically ordered component along the axis perpendicular
  to the spiral plane.

\end{abstract}

\maketitle

\section{Introduction}

The nearest-neighbor Heisenberg antiferromagnet on the kagome lattice, a
two-dimensional network of corner-sharing triangles, is one of the most
geometrically frustrated magnets. Frustration suppresses the magnetic
ordering tendency and leads to an extensive classical ground state
degeneracy.  Order-by-disorder effects lift the degeneracies in the
classical system and are believed to select the coplanar
$\sqrt{3}\times\sqrt{3}$ pattern as the ground
state.~\cite{huse1992,chubukov1992,chandra1993,reimers1993}

The spin-$1/2$ quantum kagome antiferromagnet is much less
understood. Exact
diagonalization~\cite{zeng1990,leung1993,elstner1994,lecheminant1997,waldtmann1998,sindzingre2000,misguich2007}
and series expansion studies~\cite{singh1992,misguich2005} indicate the
absence of long-range magnetic order, but the precise nature of the
ground state remains mysterious.  For small systems, numerical
simulations reveal a large number of singlet states below a small (and
possibly vanishing) spin
gap.~\cite{lecheminant1997,waldtmann1998,jiang2008} This observation has
lead to the speculation that the ground state of the kagome lattice
might be a gapless critical spin
liquid~\cite{hermele2005,ran2007,ryu2007} or a particular type of
valence-bond crystal that exhibits many low-energy singlet
states~\cite{nikolic2003,budnik2004,singh2007} 
(see also Refs.~\onlinecite{syromyatnikov2002} and \onlinecite{syromyatnikov2004}).

Recently, two new candidate materials for an ideal spin-$1/2$ kagome
antiferromagnet have attracted considerable attention. First, the
mineral ZnCu$_3$(OH)$_6$Cl$_2$, also known as Herbertsmithite, realizes
structurally undistorted, magnetically isolated kagome layers with
Cu$^{2+}$ moments on the lattice
sites.\cite{shores2005,ofer2006,helton2007,mendels2007,imai2007,bert2008,ofer2008}
Neither magnetic ordering nor spin freezing has been observed for this
material down to the lowest currently achievable temperature of 50 mK,
which is well below the energy scale of the antiferromagnetic
interaction.\cite{ofer2006,helton2007,mendels2007} While Herbertsmithite
is structurally perfect, Dzyaloshinskii-Moriya (DM) interactions and a small
number of impurities might complicate the experimental study of the
ideal quantum kagome system.\cite{rigol2007}
Secondly, there is Volborthite Cu$_3$V$_2$O$_7$(OH)$_2\cdot$2H$_2$O, a
spin-1/2 (Cu$^{2+}$) antiferromagnet, whose magnetic sublattice consists
of well-separated kagome-like
planes.~\cite{hiroi2001,fukaya2003,bert2004,bert2005} This material has
a monoclinic distortion, which deforms the equilateral kagome triangles
into isosceles triangles leading to a difference between two of the
nearest-neighbor exchange constants ($J'$) and the third one ($J$).
Similarly as in Herbertsmithite, the spins do not order down to 1.8 K, an
energy scale fifty times smaller than the exchange coupling strength in
Volborthite.~\cite{hiroi2001} However, at very low temperatures evidence
for a spin freezing transition has been reported.~\cite{bert2005} The
anisotropy ratio of the exchange couplings, $\alpha=J/J'$, could not be
determined experimentally so far, but the differing side lengths of the
kagome triangles seem to favor $\alpha > 1$. A recent comparison of
exact diagonalization calculations with thermodynamic measurements
suggests that the spatial anisotropy of the exchange couplings is small,
and that additional interactions beyond the nearest-neighbor couplings
might be present in the mineral Volborthite.~\cite{sindzingre2007}

\begin{figure}[t]
\begin{center}
\includegraphics[width=0.45\textwidth]{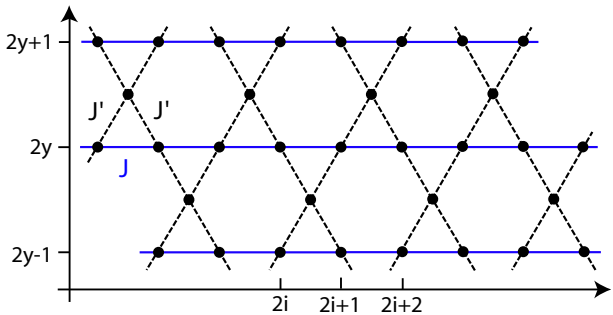}
\end{center}
\caption{
(Color online) Spatially anisotropic kagome lattice with 
nearest-neighbor exchange $J$ (blue)
among chain spins ($\bm{S}$)  
and $J'$ (black) among chain spins and interstitial spins ($\bm{s}$).
}
\label{fig: kagome lattice}
\end{figure}

In this paper, motivated by the renewed interest in the kagome systems
and the recent experiments on Volborthite, we investigate the spatially
anisotropic version of the quantum kagome antiferromagnet. We shall
focus on the quasi-one-dimensional limit, $J' \ll J$, where the model
consists of quantum-critical spin-1/2 chains weakly coupled together via
intermediate ``dangling'' spins (see Fig.~\ref{fig: kagome lattice}). In
this situation the competition between quantum fluctuations and the
strong geometric frustration of the kagome lattice is particularly
keen. The anisotropic quantum kagome antiferromagnet has been studied
previously by a variety of techniques, all of which employ perturbation
theories in some ``artificial'' small parameter.  Examples include
large-$N$ expansions of the Sp($N$) symmetric generalization of the
model,~\cite{yavorskii2007,apel2007} a block-spin perturbation approach
to the trimerized kagome lattice,~\cite{yavorskii2007} and semiclassical
calculations in the limit of large
spin.~\cite{yavorskii2007,wang2007}
Our approach is complementary to these studies in that it offers a
fairly controlled analysis of the quasi-one-dimensional limit using
powerful field-theoretical methods
\cite{starykh2004,starykh2005,starykh2007} that have originally been
developed for the investigation of quantum critical systems in one
dimension.
Recently, the quasi-one-dimensional version of the kagome
antiferromagnet in a strong magnetic field, assuming that the
intermediate spins are fully polarized, has been studied using similar
techniques.~\cite{stoudenmire2007} The present paper treats the case
of zero external field.

Our approach rests on the assumption that in the quasi-one-dimensional
limit, $J' \ll J$, the intermediate spins order at a temperature scale
$T_{{\bm{s}}}$ much higher than the ordering temperature
$T_{\mathrm{ch}}$ of the weakly coupled chains.  This is justified a
posteriori by our finding that the effective interaction among the
interstitial spins, which sets the temperature scale $T_{{\bm{s}}}$, is
of order $ (J')^2/J $, whereas the most relevant effective
interaction among the weakly coupled spin-1/2 chains, which determines
$T_{\mathrm{ch}}$, is of order $( J' )^4 / J^3 $.  Consequently,
we can divide the theoretical analysis
into three separate stages. First, we derive the effective interaction
among the intermediate dangling spins using perturbative renormalization
group (RG) transformations in the time direction. These considerations
are complemented by numerical estimates of the induced short distance
couplings among the dangling spins.  Second, we analyze the ground state
of the resulting, spatially anisotropic triangular lattice of dangling
spins as a function of first and further-neighbor interactions. We find
that in the ground state the interstitial spins form a rotating spiral
with a small (and possibly vanishing) wave vector parallel to the chain
direction.  Third we determine the most relevant interchain interactions
using a symmetry analysis and RG considerations.  Besides the effective
interchain interaction, the spiral magnetic field produced by the
intermediate spins induces another perturbation to the system of
decoupled Heisenberg chains. Finally, we analyze these perturbations to
the fixed point of the independent (decoupled) spin-1/2 chains with the
help of operator product expansions (OPEs).  

\begin{figure}[t!]
\begin{center}
\includegraphics[width=0.47\textwidth]{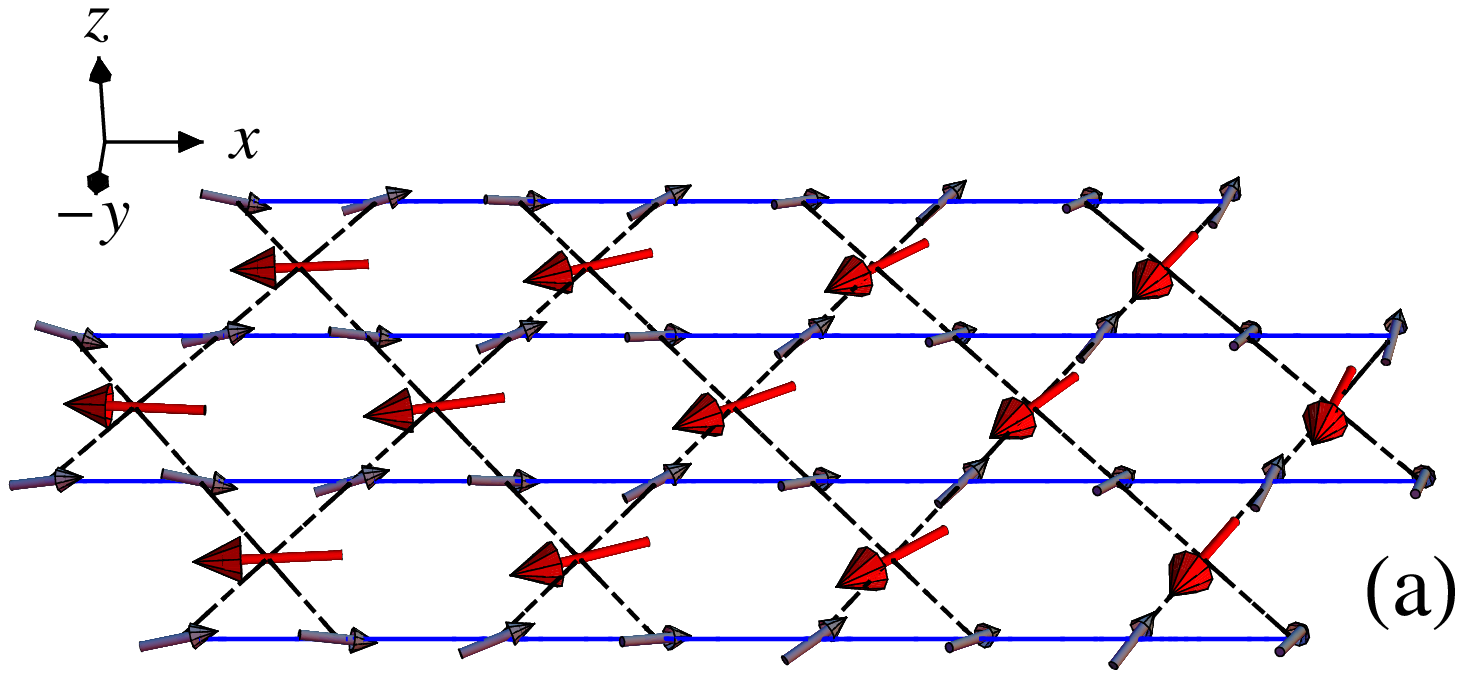}
\includegraphics[width=0.47\textwidth]{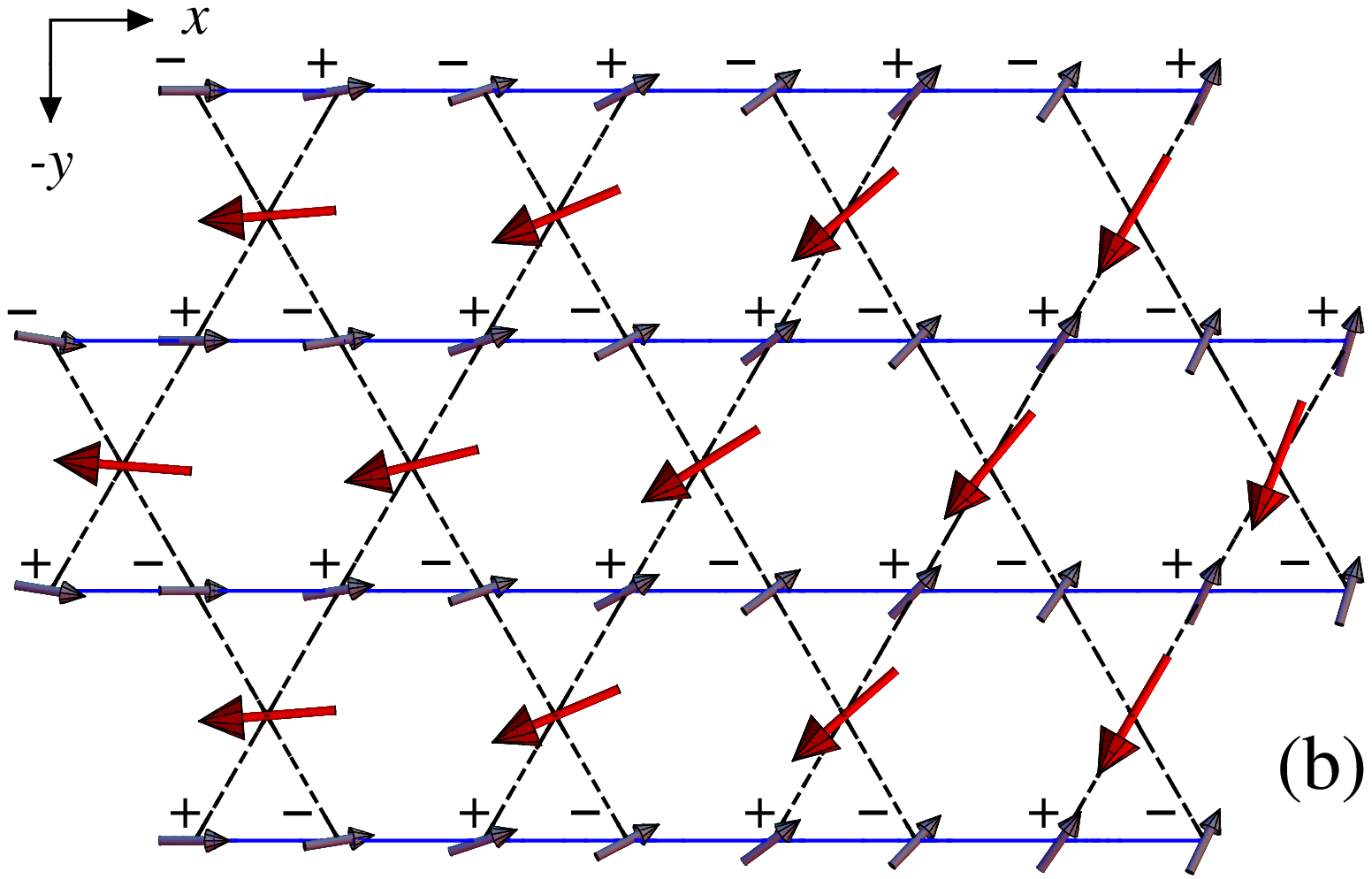}
\vspace{-0.2cm}
\end{center}
\caption{
(Color online) 
(a) Three-dimensional perspective view of the spin ordering pattern. The
interstitial spins (red arrows) form a coplanar spiral with 
an $O(1)$ local moment in the $x$-$y$ plane.
The components of the chain spins (gray arrows) in the $x$-$y$ plane
also form a spiral but with static moment of $O[J'/J]$ and antiparallel 
to the interstitial spins. The out-of-plane components ($z$ direction) of the
chain spins are non-zero but even smaller -- $O[(J'/J)^2]$ -- and
ordered antiferromagnetically along and between the chains.  
(b) Top view of the spin ordering pattern. The vertical component
of the chain spins is indicated using $+$ (upward pointing)
and $-$ (downward pointing).
}
\label{fig: kagome order}
\end{figure}

The ultimate result of our analysis is that all spins order, with the
non-coplanar configuration shown in Fig.~\ref{fig: kagome order}, in
which the interstitial and chain spins predominantly form coplanar
spirals with a wave vector $(q,0)$, but with a reduced $O(J'/J)$ static
moment on the chains.  The chain spins are weakly canted out of the
plane, with the $O[(J'/J)^2]$ normal components forming an
antiferromagnetically ordered pattern.  The precise value of $q$ cannot
be reliably determined from our analysis, but we expect $q\ll 1$ and
$q=0$ is a distinct possibility.  In the $q=0$ case, the state becomes
coplanar.  This ordered state differs from those found in two other
recent studies\cite{wang2007,yavorskii2007} using other methods, but
there are similarities.  These are discussed in Sec.~\ref{sec: conclusions}.

At first glance, it might be counter-intuitive that the interstitial 
spins and the chain spins order in a nearly ferromagnetic fashion among 
themselves rather than in an antiferromagnetic pattern. But it has to be 
kept in mind that the ordering
of the quasi-one-dimensional version of the kagome antiferromagnet 
is driven by the weakly coupled interstitial spins,
which order at a larger energy scale than the chain
spins. There is no a priori reason that the effective interaction among
the interstitial spins should be purely antiferromagnetic. In fact, it turns out
that there is an effective \emph{ferromagnetic}
interaction along the diagonal bonds connecting 
nearest neighbor interstitial spins (see Sec.~\ref{sec:numer-estim-inter}). 
Together with an effective antiferromagnetic interaction
along the horizontal bonds connecting neighboring interstitial
spins this ultimately leads to the spiral order of the dangling spins.
The spiral order of the chain spins can then be understood as arising
from the linear response to the local field of the ordered interstitial moments.

The remainder of the paper is organized as follows. Sec.~\ref{sec:
  eins} describes the lattice Hamiltonian, its symmetries, and its
low-energy field theory description. In Sec.~\ref{sec: zwei}
we analyze the low-energy interactions
using symmetry considerations and a perturbative renormalization treatment.
In Sec.~\ref{sec:numer-estim-inter} we use numerical methods
to estimate the interaction strength among the interstitial spins and
derive an effective model for the dangling spins on the triangular lattice.
Sec.~\ref{sec: vier} deals with the ground state analysis of the
triangular lattice.  The perturbative analysis of weakly coupled chains
in a spiral magnetic field with a small wave vector, $q \ll 1$, is
presented in Sec.~\ref{sec: RG} and our conclusions and a discussion of
the relation to other results are given in Sec.~\ref{sec:
  conclusions}. Some technical aspects of the RG calculations are
relegated to Appendix~\ref{app:rg-calculations}.  For completeness, we
study in Appendix~\ref{sec:large-q} the ordering of the weakly coupled
chains in the presence of a spiral field with a large wave vector.

\section{Model Definition and low-energy Hamiltonian}
\label{sec: eins} 

The Hamiltonian of the Heisenberg antiferromagnet 
on the anisotropic kagome lattice (see Fig.~\ref{fig: kagome lattice})
is given by $H=H_0 + V$, where $H_0$
describes the decoupled set of chains with nearest-neighbor antiferromagnetic Heisenberg interactions
\begin{subequations} \label{lattice ham}
\begin{eqnarray} \label{lattice ham a}
H_0
=
J \sum_{i,y} \bm{S}_{i,y} \cdot \bm{S}_{i+1,y},
\end{eqnarray}
and $V$ is the interaction
among the chains and the intermediate spins 
\begin{eqnarray}
\label{def V}
V
&=&
J'
\sum_{i,y} \bm{s}_{2i \pm \frac{1}{2}, 2y \mp \frac{1}{2} } \cdot
\Big(
\bm{S}_{2i,2y} + \bm{S}_{2i \pm 1, 2y} 
\nonumber\\
&& \qquad \qquad
+ \bm{S}_{2i, 2y \mp 1} + \bm{S}_{2i \pm 1, 2y \mp 1} 
\Big) .
\end{eqnarray}
\end{subequations}
For the interstitial spins we shall use the symbol $\bm{s}$, whereas the
chain spins are denoted by $\bm{S}$.
The anisotropic kagome lattice has rotation, translation and reflection
symmetries.  The translational subgroup is generated by the translations
$T_1$ and $T_2$, which move the lattice by two units along the
horizontal, and one of the diagonal axes, respectively. The rotational
subgroup consists of $\pi$ rotations about the lattice sites and about
the centers of the hexagons.  We can distinguish two types of reflection
symmetries: reflections $R_1$ about a vertical line through a midpoint
of a chain bond (link parity), and reflections $R_2$ about a horizontal
line passing through the intermediate dangling spins (see Fig.~\ref{fig:
  kagome lattice}).
 
Let us now describe the chains in the continuum limit, which is
applicable as long as $J \gg J'$.  In this limit the low-energy
properties of the antiferromagnetic spin-1/2 chains are governed by the
Wess-Zumino-Novikov-Witten (WZNW) SU(2)$_1$ theory, and the chain spin operator
$\bm{S}_{i,y}$ can be decomposed into its uniform $\bm{M}_y(x)$ and
staggered $\bm{N}_y(x)$ spin magnetizations 
\begin{subequations} \label{continuum fields}
\begin{eqnarray} \label{S in continuum}
\bm{S}_{i,y} 
\rightarrow
a_0 \left[ \bm{M}_y (x)  + (-1)^x \bm{N}_y (x) \right] ,
\end{eqnarray}
where $x=i a_0$ and $a_0$ denotes the lattice spacing.
The uniform magnetization can be written in terms of left- and right-moving SU(2) currents,
$\bm{M}_y = \bm{J}_{y,R} + \bm{J}_{y,L}$. 
Another important operator describing low energy properties of
the spin-1/2 chains is the staggered dimerization $\varepsilon_y (x)$, 
which is defined as the 
continuum limit of
the scalar product of two neighboring spins
\begin{eqnarray}
\bm{S}_{i,y}   \cdot \bm{S}_{i+1,y}  
\rightarrow (-1)^x \varepsilon_y(x) .
\end{eqnarray}
\end{subequations}
The scaling dimension of these continuum operators determines the
relevance of the operator in the RG sense with respect to the Luttinger
liquid fixed point of the decoupled chains.  The uniform magnetization
$\bm{M}_y$ has scaling dimension $1$, whereas both the staggered spin
magnetization $\bm{N}_y $ and the staggered dimerization $\varepsilon_y$
have scaling dimension $1/2$.
The microscopic lattice symmetries of $H$, Eq.~\re{lattice ham}, leave
an imprint on continuum description~\re{continuum fields}.  That is,
the action of the space group symmetries on the continuum operators is
given by
\begin{subequations} \label{eq: symmetries}
  \begin{eqnarray}
T_1 &:&  
\quad
\bm{M} \to \bm{M},
 \quad
 \bm{N} \to  + \bm{N},
 \quad
\varepsilon \to   +\varepsilon,
\\
R_1 &:& 
\quad
 \bm{M} \to \bm{M},
 \quad
 \bm{N} \to - \bm{N},
 \quad
\varepsilon \to  +\varepsilon,
\\
 \label{eq: third symmetry}
R_2 \circ T_2 
&:& 
\quad
  \bm{M} \to \bm{M},
 \quad
 \bm{N} \to - \bm{N},
 \quad
\varepsilon \to - \varepsilon ,
 \end{eqnarray}
\end{subequations}
where, for brevity, we have suppressed the chain index.  Other symmetry
operations on the continuum fields are either trivial, or can be
rewritten as a product of the above transformations.

The three continuum fields $\bm{M}$, $\bm{N}$, and $\varepsilon$ form a closed operator algebra
with respect to certain operator product expansions, which are widely used in 
the literature.\cite{starykh2004,starykh2005,starykh2007,senechal1999,delft98,cardy96,lin97} For example, the right-moving SU(2) currents
$\bm{J}_R$ satisfy the following chiral OPEs
\begin{subequations} \label{eq: OPEs}
\begin{eqnarray} \label{eq: OPEsa}
J^a_{R} ( x, \tau ) J^b_{R} (0 )
&=&
\frac{ \delta^{ab} / ( 8 \pi^2 ) }{   ( u\tau -ix + a_0 \sigma_{\tau}  )^2 }
+
\frac{ i \varepsilon^{abc} J^c_{R} (0 ) / ( 2 \pi ) }{ u \tau -ix +a_0 \sigma_{\tau}  }, 
\nonumber\\
\end{eqnarray}
with imaginary time $\tau$, $\sigma_{\tau} = \mathrm{sgn} \, \tau$, the short-distance
cut-off $a_0$, and spin velocity $u = \pi J a_0/2$.
Similar relations hold for the left-moving spin currents $\bm{J}_L$.
The product of  $\bm{J}_R$ and $\bm{N}$ can be expanded as
\begin{eqnarray} \label{ope JN}
J^a_R ( x, \tau ) N^b (0)
&=&
\frac{ i \epsilon^{abc} N^c (0) - i \delta^{ab} \epsilon (0) }
{ 4 \pi ( u\tau - ix + a_0 \sigma_{\tau} ) } .
\end{eqnarray}
\end{subequations}
The above equalities are understood to be valid only when inserted into
correlation functions and in the limit where the two points $(x, \tau)$
and $(0,0)$ are close together. These current algebra relations will
allow us to compute one-loop RG equations by purely algebraic means (see
Secs.~\ref{sec: zwei} and \ref{sec: RG}).

Using relation~\re{S in continuum}, we can derive a naive continuum
limit of the interaction among the chains and the intermediate spins,
Eq.~\re{def V}.  First, we note that the intermediate spins
$\bm{s}_{2i+\frac{1}{2},2y-\frac{1}{2}}$ couple symmetrically to
$\bm{S}_{2i,2y}$ and $\bm{S}_{2i+1,2y}$. Hence, the staggered spin
magnetization enters only via its first derivative in the continuum
version of the interaction $V$, and we can set
\begin{eqnarray}~\label{cont lim two spin}
\bm{S}_{2i,2y} +
\bm{S}_{2i+1,2y} 
\rightarrow
2 a_0 \bm{M}_{2y} (2 x) -  \frac{a_0^2}{2} \partial_x \bm{N}_{2y} (2 x).
\end{eqnarray}
In the above, and the following, we use the notation that the derivative
in $\partial_x \bm{N}_{2y}(2x)$ is with respect to the full argument,
i.e., more explicitly 
\begin{equation}
  \label{eq:7}
  \partial_x \bm{N}_{2y}(2x) \equiv \left.\partial_X \bm{N}_{2y}(X)\right|_{X=2x}.
\end{equation}
With this, the interaction Eq.~\re{def V} reads $V=V_1 + V_2$,
with
\begin{eqnarray} \label{cont int}
V_1
&=&
\gamma_1 \sum_{x,y }  
  \bm{s}_{2x \pm \frac{1}{2}, 2y \mp \frac{1}{2} } \cdot
\left[  \bm{M}_{2y} ( 2x ) + \bm{M}_{2y \mp 1} ( 2x ) \right]   ,
\nonumber\\
\\
V_2
&=& 
 \gamma_2 \sum_{x,y }  
(\pm )
  \bm{s}_{2x \pm \frac{1}{2}, 2y \mp  \frac{1}{2} } \cdot
 \partial_x \left[
 \bm{N}_{2y} (2x ) 
+    \bm{N}_{2y \mp 1} (2x) 
\right],
\nonumber
\end{eqnarray}
where the bare coupling constants are given by, $\gamma_1 = 2 J' a_0$
and $\gamma_2 = - J' a_0^2/2$, which follows from substituting \re{cont
  lim two spin} into~\re{def V}.  We have retained the next-to-leading
interaction $V_2$ as it will produce, in combination with the leading
term $V_1$, relevant interchain couplings with respect to the fixed point
of the decoupled chains (see Sec.~\ref{sec: zwei}).  
The continuum description of the interaction $V$, Eq.~\re{cont int}, is
necessarily invariant under the symmetry transformations, Eq.~\re{eq:
  symmetries}.  In particular, we note that $V$ does not contain a
contribution $\bm{s}_{2x+\frac{1}{2},2y-\frac{1}{2} }\cdot \bm{N}_{2y} (2x)
$ which is forbidden by the symmetries of the spatially anisotropic
kagome lattice (link parity $R_1$ sends $\bm{N} \to -\bm{N}$).

\section{Renormalization Group Treatment}
\label{sec: zwei}

Here, we study the effective low-energy interactions among the dangling
spins ($H_{\triangle}$), as well as those between nearest neighbor chains
($V_{\mathrm{ch}}$), using both a symmetry analysis and perturbative
renormalization group transformations.  The goal is to understand which
interactions amongst interstitial spins, amongst chains, or betwixt the
two, are most relevant, and further, at what energy scales they become
important in determining the low-energy physics.  
Technically, the Hamiltonian \re{lattice ham} with the chains treated in the
continuum limit is formally rather similar to a Kondo lattice. As the
Hamiltonian retains some local character due to the intermediate spins,
in the framework of the RG approach, this problem develops only in time,
and the energy is the only variable that is being rescaled as the RG
progresses.  Such an RG scheme is similar in spirit to the one employed
in the context of a single impurity coupled to a Luttinger liquid (see,
for example, Refs.~\onlinecite{affleck1991,gogolin1998,eggert2001}).

 \subsection{Symmetry analysis}

 Before proceeding with the RG derivation of the low-energy effective
 interactions, we first write down the most general form of
 $V_{\mathrm{ch}}$ and $H_{\triangle}$ that are allowed by the
 symmetries of the spatially anisotropic kagome lattice.  Such a general
 symmetry consideration will reveal all relevant symmetry allowed
 interactions that are expected to be generated through RG
 transformations, when all nominally irrelevant terms are taken into
 account.  The space group symmetries needed for this analysis have been
 discussed in Sec.~\ref{sec: eins} [see Eqs.~\re{eq: symmetries}].
 
 We begin with the interchain Hamiltonian $V_{\mathrm{ch}}$.  In
 principle, the number of allowed interchain interaction terms is
 infinite. However, only a handful of terms will be important.  Most
 significant are those terms which are most relevant in the RG sense
 (with respect to the Luttinger liquid fixed point of the decoupled
 chains).  This amounts to two-chain interactions involving no
 derivatives and only continuum fields that have small scaling
 dimension.  Amongst these, we may further restrict ourselves to
 nearest-neighbor chain interactions, as the magnitude of
 further-neighbor chain interactions is expected to decrease with
 separation distance.  The continuum operators with the smallest scaling
 dimension are the staggered dimerization $\varepsilon_y$ and the
 staggered magnetization $\bm{N}_y$.  Therefore, we find
\begin{eqnarray} \label{eq: interchain}
V_{\mathrm{ch}}
=
\sum_y
\int dx 
\left\{
\gamma_N \bm{N}_y \cdot \bm{N}_{y+1}
+
\gamma_{\varepsilon} \varepsilon_y \varepsilon_{y+1}
\right\} ,
\end{eqnarray}
where the value of the coupling constants $\gamma_N$ and
$\gamma_{\varepsilon}$ will have to be determined by microscopic
calculations. Using Eqs.~\re{eq: symmetries} it is straightforward to
check that these are the only terms with lowest possible scaling
dimension $1$ that satisfy the symmetry requirements of the spatially
anisotropic kagome lattice.  

In addition to these most relevant interchain interactions, it is
necessary to consider also a few less relevant terms, as these will
arise at lower order in the renormalization group treatment below, and
are important for {\sl generating} the more relevant terms in
Eq.~(\ref{eq: interchain}) above.  These are
\begin{equation}
  \label{eq:1}
  V_{\mathrm{ch}}^{(1)}
=
\sum_y
\int dx 
\left\{
\gamma_{\scriptscriptstyle\partial N} \partial_x \bm{N}_y \cdot \partial_x\bm{N}_{y+1}
+
\gamma_M \bm{M}_y \cdot \bm{M}_{y+1}
\right\} .
\end{equation}

\begin{figure}[t]
\begin{center}
\includegraphics[width=0.38\textwidth]{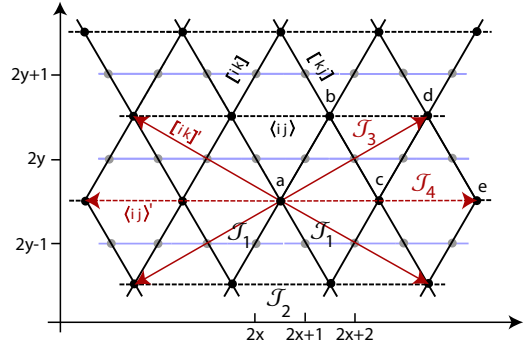}
\end{center}
\caption{
(Color online) Exchange paths of the triangular lattice formed by 
the intermediate spins $\bm{s}$ of the 
spatially anisotropic kagome lattice. Solid black lines denote the nearest-neighbor ferromagnetic 
exchange $\mathcal{J}_1$ (a - b and b - c bonds), dotted black lines the nearest-neighbor antiferromagnetic exchange $\mathcal{J}_2$ (a - c bond).
The further neighbor interactions $\mathcal{J}_3$ (a - d bond)
and $\mathcal{J}_4$ (a - e bond) for the spin $\bm{s}_a$ 
are illustrated with red arrows. The chain bonds of the anisotropic
kagome lattice are indicated in light blue (cf.~Fig.~\ref{fig: kagome lattice}).
}
\label{fig: triangular lattice}
\end{figure}

Next, we turn to interactions among the interchain spins $H_{\triangle}$.
Considering only terms that arise at $O[(J')^2]$, we find that $H_{\triangle}$ is given
by Heisenberg interactions among dangling spins whose $y$ coordinates differ by at most one unit.
 The coupling constants of these Heisenberg interactions are restricted by the symmetries
of the lattice. With these conditions, $H_{\triangle}$ can be conveniently written in the form (see Fig.~\ref{fig: triangular lattice})
\begin{eqnarray} \label{H dreieck}
H_{\triangle}
&=&
\sum_{x,y, r > 0}  
{\mathcal J}_{2r}  \,
\bm{s}_{2x \pm \frac{1}{2}, 2y \mp \frac{1}{2} } \cdot \bm{s}_{2x \pm \frac{1}{2} + 2 r, 2y \mp \frac{1}{2} }
\\
&& \qquad
+
 \sum_{x,y,r}  {\mathcal J}_{| 2 r +1 |} 
 \bm{s}_{2x \pm \frac{1}{2}, 2y \mp \frac{1}{2} } \cdot \bm{s}_{2x \pm \frac{3}{2} + 2r, 2y \pm \frac{1}{2} } .
 \nonumber
\end{eqnarray}
Since the couplings $J_r$ decrease in magnitude with increasing $|r|$,
we can truncate the above sums over $r$ after the first few terms.

\subsection{Renormalization group results}
\label{subsec:rg}

To determine the fluctuation-generated corrections to the low-energy
effective action we perform a perturbative RG analysis of the
interactions $V$, Eq.~\re{cont int}, to one-loop order.  The perturbation
theory is formulated by expanding the partition function $
Z = \int e^{-S_0- \int d \tau V} $
up to quadratic order in the couplings
\begin{eqnarray}  \label{eq: pert exp}
Z &\simeq &
\int e^{-S_0} \left[
1 - \int d \tau V
+
\frac{1}{2} \mathrm{T} \hspace{-0.1cm}
\int d \tau_1 d \tau_2
V ( \tau_1 ) V ( \tau_2 )
\right]  .
\nonumber\\
\end{eqnarray}
Here,  $S_0$ denotes the fixed point action,
$\mathrm{T}$ is the time-ordering operator, and $\tau_i$ is the imaginary
time.  Implicit in Eq.~(\ref{eq: pert exp}) is a regularization, such that the integrals appearing in
the expansion are taken only over the regions in which no two times are
closer than  some short time cut-off $\alpha = a_0/u$.  The RG proceeds by
increasing this cut-off infinitesimally from $\alpha$ to $b \alpha$,
where $b= e^{d\ell}$ and $d\ell>0$ is the usual logarithmic change of
scale.  To do this, all pairs of times $\tau_1,\tau_2$ such that
$\alpha<|\tau_1-\tau_2|< b\alpha$ must be fused using the
operator product expansion, and the integral over $\tau_1-\tau_2$ in
this range carried out.  One thereby obtains a new partition function
with renormalized interactions and the increased cut-off.  We then
perform an additional trivial rescaling of time and space coordinates
for to restore the original cut-off:
\begin{eqnarray}      \label{eq:5}
  {\bf N}_y(x,\tau) & \rightarrow & b^{-1/2} {\bf N}_y(x/b,\tau/b), \nonumber\\
  {\bf M}_y(x,\tau) & \rightarrow & b^{-1} {\bf M}_y(x/b,\tau/b), \\
  {\bf s}_{x,y}(\tau) & \rightarrow & {\bf s}_{x,y}(\tau/b). \nonumber
\end{eqnarray}
We note that the spatial rescaling is not required to restore the
cut-off but is natural and convenient for the continuum fields of the
chains, which are described by a Lorentz invariant field theory.  By
contrast, we obviously cannot rescale the coordinates of the discrete
spin operators of the interstitial spins.
This difference leads to the appearance of an explicit RG length scale
in the effective couplings between the chains and interstitial spins,
$V_1$ and $V_2$, in the renormalized Hamiltonian:
\begin{eqnarray}  \label{eq:6}
  V_1(\ell)
&=&
\gamma_1 \sum_{x,y }  
  \bm{s}_{2x \pm \frac{1}{2}, 2y \mp \frac{1}{2} } \cdot
\left[  \bm{M}_{2y} ( 2x_\ell ) + \bm{M}_{2y \mp 1} ( 2x_\ell) \right]   ,
\nonumber\\
V_2(\ell)
&=& 
 \gamma_2 \sum_{x,y }  
(\pm )
  \bm{s}_{2x \pm \frac{1}{2}, 2y \mp  \frac{1}{2} } \cdot
\partial_x  \big[  
 \bm{N}_{2y} (2x_\ell)  
 \\
&& \qquad \qquad \qquad  \qquad  \qquad  \qquad
+     \bm{N}_{2y \mp 1} (2x_\ell) 
\big],
\nonumber
\end{eqnarray}
where $x_\ell= xe^{-\ell}$.

With this, the derivation of the RG equations is straightforward, and follows closely the methods used
in Refs.~\onlinecite{affleck1991,gogolin1998,eggert2001}.  Details of
the calculations can be found in Appendix~\ref{app:rg-calculations}.  Taking into account
$V=V_1+V_2$ to one loop, we obtain  RG equations for both the
interchain couplings and the couplings between the interstitial spins.
The former are
\begin{eqnarray} \label{eq:4}
 \frac{ d \gamma_1 }{ d \ell }
&=&
+\frac{1}{\pi u} \gamma_1^2,
\qquad \quad \quad \; \; \,
\frac{ d \gamma_2 }{ d \ell}
= - \frac{1}{2} \gamma_2
 +\frac{1}{\pi u} \gamma_1 \gamma_2 ,
\nonumber\\
\frac{ d \gamma_{\scriptscriptstyle \partial N}}{ d \ell }
&= & - \gamma_{\scriptscriptstyle \partial N}
- \frac{\gamma_2^2}{u}, 
\qquad \; \; 
\frac{ d \gamma_M }{ d \ell}
= 
- \gamma_1^2 + \frac{\gamma_M^2}{2\pi u},
\\
\frac{d \gamma_N}{d\ell} & = & \gamma_N + 
\frac{\gamma_{\scriptscriptstyle \partial N} \gamma_M}{8\pi\alpha^2 u}
, \qquad
\frac{d \gamma_\varepsilon}{d\ell} = \gamma_\varepsilon -
\frac{3\gamma_{\scriptscriptstyle \partial N} \gamma_M}{16\pi\alpha^2 u} .\nonumber
 \end{eqnarray}
 Note that the strongly relevant interactions $\gamma_N,\gamma_\epsilon$
 are {\sl not} generated at one loop from $\gamma_1,\gamma_2$.  Instead,
 the former are generated from the sub-dominant
 $\gamma_{\scriptscriptstyle \partial N},\gamma_M$ terms.  This is the
 reason the latter terms needed to be included in our treatment.

The RG equations for the interstitial spin couplings are {\sl
  functional}, insofar as they describe the flow of the full set of
interactions ${\mathcal J}_r$.  We find
\begin{subequations}   \label{eq:2}
\begin{eqnarray}
  \frac{ d {\mathcal J}_{2r}}{ d \ell }
&=& {\mathcal J}_{2r}+ I_M(r e^{-\ell}) \frac{\gamma_1^2}{u} + 
I_N(r e^{-\ell}) \frac{\gamma_2^2}{u},
\end{eqnarray}
where
\begin{eqnarray}
  \label{eq:3}
  I_M(r) & = & \sum_{\sigma'=\pm 1} \frac{\alpha}{4 \pi^2} \frac{1}{
    \left[ \alpha +\sigma' i 2 r \right]^2 },
  \nonumber \\
I_N(r) & = & 8 C_N \frac{ \alpha(\alpha^2 - 8 r^2)}{( \alpha^2 + 4 r^2)^{5/2}} ,
\end{eqnarray}
\end{subequations}
where $C_N$ is the amplitude of the $\langle N^a N^a \rangle$
correlator. Analogous expressions can be derived for the interactions
$\mathcal{J}_{|2r+1|}$, i.e., the second term in $H_\triangle$,
Eq.~\re{H dreieck}.

In passing, we note that the form of the RG flows in Eqs.~(\ref{eq:4})
is highly constrained by the symmetry of the full (spatially
anisotropic) kagome lattice, and this is necessary to avoid the
generation of strongly relevant interactions at this order.  By
contrast, for the kagome {\sl strip} of extension one in the 
$y$ direction, there is no translational symmetry $T_2$, and consequently
symmetry \re{eq: third symmetry} is absent.  This allows for the
appearance of a term in the renormalized Hamiltonian proportional to 
$\int d x  \varepsilon_y (x, \tau)$, which is indeed generated
proportionally to $\gamma_1\gamma_2$ in that case (compare with \ Azaria
\textit{et al.}, Ref.~\onlinecite{azaria1998}, and see Ref.~\onlinecite{white00}).  
This term is strongly
relevant and generates a gapped dimerized state in that model.

\subsection{Implications}
\label{sec:implications}

The RG flows in Eqs.~(\ref{eq:4},\ref{eq:2}) give considerable insight
into the emergent energy scales in the problem.  Though interactions
between the chains and between the interstitial spins are both generated
from the bare interactions $\gamma_1,\gamma_2$, their characters are
distinctly different.  Crucially, we see from Eqs.~(\ref{eq:4}) that
amongst the chains, only marginal and irrelevant interactions --
$\gamma_{\scriptscriptstyle \partial N}$, $\gamma_M$ -- are generated at
this order.  Since the bare values $\gamma_1(\ell=0) \sim
\gamma_2(\ell=0)$ are both of $O[J']$, this includes all effective
generated interactions up to $O[(J')^2]$.  A simple scaling argument
shows that the low-energy excitations of the chains can be modified from
those of decoupled chains only on energy scales {\sl parametrically
  smaller} than $(J')^2$.  

To understand this, let us consider the solutions to
Eqs.~(\ref{eq:4}) in somewhat more detail.  The most important
observation is that the strongly relevant interactions,
$\gamma_N,\gamma_\varepsilon$ [third line of Eqs.~(\ref{eq:4})], once
they are generated, grow rapidly and exponentially with $\ell$.  They
are therefore dominated by the effects of their ``source'' terms
(proportional to $\gamma_{\scriptscriptstyle \partial N}\gamma_M$) at
{\sl small} $\ell$, which are most amplified under the growth.  The
source terms at small scale are themselves of order $(J')^2/J$, as
described above.  Therefore we expect that the relevant interactions
will be, at least until they renormalize to large values, of order
$(J')^4/J^3$.  Indeed, the ultimate effect of all this short-scale
renormalization is completely equivalent to including
fluctuation-generated ``bare'' interactions of this same order.  This
procedure has been described in considerable detail in prior
publications, in which the formal manipulations closely parallel those
used here.  See Refs.~\onlinecite{starykh2004,starykh2007} for further information.
Therefore we will in Sec.~\ref{sec: RG} simply take the relevant
couplings to have the appropriate values:
\begin{eqnarray}
\gamma_N(\ell \sim 1)
=
+ \frac{ 2 ( J' )^4 }{ 4 \pi^4 J^3 } ,
\qquad
\gamma_\varepsilon(\ell \sim 1)
=
-  \frac{ 3 ( J' )^4 }{ 4 \pi^4 J^3 }.
\label{ap:relevant}
\end{eqnarray}

By contrast, the interactions ${\mathcal J}_r$ generated in
Eqs.~(\ref{eq:2}) at $O[(J')^2]$ amongst the dangling spins will clearly
modify their energetics at this same order, since they act within an
otherwise completely degenerate manifold.  More formally, this follows
directly from the RG flows, Eq.~(\ref{eq:2}), which shows that the
interstitial interactions will grow to $O(1)$ values (actually to any
fixed value) by a scale $Je^\ell \sim J (\gamma_1^2,\gamma_2^2) \sim
(J')^2/J$.  Therefore we  see that, as
claimed in the introduction, the ordering (as we shall see, this is the
result of these interactions) of the interstitial spins occurs at an
energy scale at which the chains are unaffected by their couplings to
the interstitial spins and each other.

Having understood the hierarchy of energy scales resulting from the RG,
we can proceed to try to understand in more detail the ordering of the
interstitial spins.  In principle, one may  do so by
integrating the flow equations for the ${\mathcal J}_r$ couplings,
Eqs.~(\ref{eq:2}), from $\ell=0$ to the ordering scale, $\ell=2 \ln
(J/J')$.  This is equivalent to performing the time integrals not over
the infinitesimal shell in the vicinity of the running cut-off, but
instead from microscopic times $\alpha$ up to times of order $\alpha
(J/J')^2$.  Inspection of expressions in Eq.~(\ref{eq:3}) shows that
these integrals in fact are dominated by times of order the microscopic
time. Indeed the upper limit of the time integration can be extended
to infinity without significant modification of the result. 

The dominance of short times has physical significance: it implies that
the generated interstitial spin interactions are in fact controlled by
high-energy physics of the Heisenberg chains, i.e., correlations of the
chain spins on the scale of a few lattice spacings.  These short
distance correlations are {\sl not} precisely captured by the continuum
bosonization/current algebra approach.  Therefore, if an accurate
determination of these dangling spin interactions is desired, we must
abandon (for the moment) the field theory methodology and return to a
direct study of the lattice model.  Conversely, because the generated
interactions are dominated by short distance physics, we expect that no
serious divergences will arise in a microscopic calculation.  In the
following section we show how perturbative and numerical methods may be
used to perform the necessary calculations and obtain the fluctuation
generated interactions amongst the interstitial spins.  Using these
precise results we determine the interstitial spin ordering in
Sec.~\ref{sec: vier}.  

Following this, in Sec.~\ref{sec: RG} we will return to the lower energy
physics resulting from the fluctuation-generated interactions
$\gamma_{\scriptscriptstyle \partial N}, \gamma_M$ in Eqs.~(\ref{eq:4}),
which will ultimately lead to ordering amongst the chain spins as well.

\section{Numerical estimates of Interactions among dangling spins}
\label{sec:numer-estim-inter}

The dynamical structure factor 
of the spin-1/2 antiferromagnetic Heisenberg chain,
$S(q, \omega)$, can be related to the coupling constants  of the effective
interactions among the dangling spins, $\mathcal{J}_r$, by means of second-order 
degenerate perturbation theory in the interaction $V$, Eq.~\re{def V}.
Using available numerical data for $S(q, \omega)$,
both in the two-spinon approximation and for finite-size systems,
this gives an estimate for the sign and relative strength of the interactions $\mathcal{J}_r$.

To formulate the perturbation theory in $V$ we denote the ground state 
of the unperturbed Hamiltonian $H_0$, Eq.~\re{lattice ham a}, 
by $|0\rangle$ and
introduce the ground-state projection operator $P=|0\rangle\langle 0|$.
With $Q=1-P$, any given state $|\Psi\rangle$ of chain ($\bm{S}$) and interchain ($\bm{s}$)
spins can be written as $|\Psi\rangle = 
|\Psi_0\rangle + |\Psi_Q\rangle$, where
$ |\Psi_0\rangle =  P |\Psi\rangle$ and $ |\Psi_Q\rangle = Q |\Psi\rangle$.
We note that $|\Psi_0\rangle$ describes the state with the chain spins in the ground state,
while the state of the $\bm{s}$-spins remains arbitrary, i.e., 
$|\Psi_0\rangle = |0\rangle |\{\bm{s}\}\rangle$. 
Following along the lines of Refs.~\onlinecite{bohm1993,stoudenmire2007},
we obtain an eigenvalue equation for $|\Psi_0\rangle$ 
\begin{equation}\label{eq:dpt4}
\Big( H_0 + P V \frac{1}{1-R Q V} R V \Big) |\Psi_0\rangle = E |\Psi_0\rangle ,
\end{equation}
with the resolvent $R = (E - H_0)^{-1}$.
Eq.~\re{eq:dpt4} is highly non-linear in the energy $E$, 
since the left-hand side of this equation depends on $E$ through the resolvent $R$.
Performing a perturbation expansion on the operator $(1-RQV)^{-1}$ gives
at second order in $J'$  
\begin{equation}
\Big( H_0 + P V R_0 V \Big) |\Psi_0\rangle = E |\Psi_0\rangle ,
\label{eq:dpt5}
\end{equation}
where $R_0 = (E_0 - H_0)^{-1}$ is a function of the ground state 
energy $E_0$ of the decoupled chains.
Multiplying Eq.~\re{eq:dpt5} from the left by $\langle 0|$ 
and inserting the perturbation $V$, Eq.~\re{def V}, yields,
after some algebra,
a Schr\"odinger equation for the interchain spins alone
\begin{subequations} \label{eq: dpt}
\begin{eqnarray}
&& 
H_{\triangle} \left|\{\bm{s}\} \right\rangle 
=
(E - E_0) \left|\{\bm{s}\} \right\rangle ,
\end{eqnarray}
with the effective interaction among interstitial spins
\begin{eqnarray}  \label{eq:dpt6}
&&
\hspace{-0.4cm}
H_{\triangle}
=
4 (J')^2  \hspace{-0.25cm}  \sum_{x,y, r > 0}
A ( 2 r) 
\left[
 \bm{s}_{2x \pm \frac{1}{2},2y \mp \frac{1}{2}}  \cdot
\bm{s}_{2x \pm \frac{1}{2} + 2r,2y \mp \frac{1}{2}}  
\right]
\nonumber\\
&&
\hspace{-0.15cm}
+2(J')^2  \hspace{-0.1cm} \sum_{x,y, r  }
A ( | 2r +1 | )
\left[
 \bm{s}_{2x \pm \frac{1}{2},2y \mp \frac{1}{2}}  \cdot
\bm{s}_{2x \pm \frac{3}{2} + 2r,2y \pm \frac{1}{2}} 
\right] .
\nonumber\\
\end{eqnarray}
\end{subequations}
Here, $A(r)$ is the linear combination 
\begin{eqnarray} \label{eq: lin. comb}
A(r)
=
G_M (r-1 )+2 G_M (r ) + G_M( r+ 1) 
\end{eqnarray}
 of the ground state expectation values
\begin{eqnarray} \label{eq: GM}
G_M  (r)
&=&
 \langle 0 | S^a_{2x,y} \frac{1}{E_0 - H_0} S^a_{2x+r,y}  |0\rangle .
\end{eqnarray}
In deriving equation~\re{eq: dpt} we made use of the fact that
the expectation value  \re{eq: GM} is independent of both the $x$
and $y$ coordinates, due to translational invariance.
The linear combination in Eq.~\re{eq: lin. comb} arises, because
every interstitial spin $\bm{s}$ is coupled to pairs of chain spins $\bm{S}$.
From inspection of Eq.~\re{eq: dpt} we find the following expressions for the exchange couplings 
$\mathcal{J}_r$, which have been defined in Eq.~\re{H dreieck},
\begin{eqnarray} \label{Jr w dpt}
\mathcal{J}_{2r}
&=&
4 ( J' )^2 A ( 2 r ) ,
\nonumber\\
\mathcal{J}_{ 2r+1}
&=&
2 ( J' )^2 A ( 2 r +1 ) ,
\end{eqnarray}
with $r > 0$. 
The factor of $2$ difference between the first
and the second lines in Eq.~\re{Jr w dpt}
accounts for the fact that interstitial spins $\bm{s}$ with
the same $y$ coordinate, say, $y-\frac{1}{2}$,  are connected 
via both the $y$ and $y-1$ chains. 

Finally, by inserting a resolution of identity in Eq.~\re{eq: GM}, 
\begin{eqnarray} \label{eq:dpt7}
G_M(r)
= 
\sum_{n \neq 0} \langle 0 | S^a_{2x,y}| n \rangle
\frac{1}{E_0 - E_n} \langle n | S^a_{2x+r,y}  |0\rangle \nonumber ,
\end{eqnarray}
with $| n \rangle$ denoting the eigenstates of $H_0$,
we realize that $G_M (r)$, Eq.~\re{eq: GM}, is just a spectral representation
of the zero-frequency Matsubara spin Green's function 
\begin{eqnarray}
G_M (r) 
= 
- \int_0^{\infty } d \tau \langle \mathrm{T}_{\tau} S^a (2x+r, \tau ) S^a  (2x,0 ) \rangle.
\end{eqnarray}
The zero-frequency Green's function $G_M (r)$ in turn is connected  to the
dynamical structure factor $S(q, \omega)$ via a Kramers-Kronig transform
 \begin{eqnarray}
 G_M (r)
&=& 
  \frac{2}{\pi}  
\int_0^\infty d \omega' \int_{0}^{+ \pi} d q \,
S (q, \omega') 
 \frac{\cos ( q r )}
 { \omega^{\prime }} .
\end{eqnarray}
Inserting the above equation into the definition 
of  $A(r)$, Eg.~\re{eq: lin. comb}, gives
\begin{eqnarray}
\label{eq: master formula Al}
A(r)
&=&
  \frac{8}{ \pi}  
\int_0^\infty d \omega' \int_{0}^{ \pi} d q 
\cos^2  \frac{q}{2} 
S  (q, \omega') 
 \frac{\cos ( q r )}
 { \omega^{\prime }} .
\end{eqnarray} 

Having related the interstitial exchange interactions $\mathcal{J}_r$ to the dynamical 
structure factor, we evaluate numerically $S(q,w)$ in the subsequent sections
to obtain precise estimates for the first few interstitial couplings  $\mathcal{J}_1 \cdots \mathcal{J}_4$.

\subsection{Two-spinon dynamical structure factor}

First we compute the couplings $\mathcal{J}_r$ using
the two-spinon approximation for $S(q, \omega)$.
 The two-spinon contribution to 
 the dynamical structure factor
can be explicitly written as (see Refs.~\onlinecite{karbach1997} and~\onlinecite{caux2006})
\begin{eqnarray}
S_2 (q, \omega)
=
\frac{1}{2 \pi } \frac{e^{-I( \rho ( q, \omega))}}{ \sqrt{ \omega^2_{U} (q) - \omega^2}}
\Theta[ \omega^{U}_q - \omega ] \Theta [ \omega^{L}_q - \omega ]  ,
 \end{eqnarray}
with the fundamental integral
\begin{eqnarray} \label{aux integral}
I(\rho)
=
\int_0^{\infty} 
d t \, \frac{ e^t}{ t} 
\frac{ \cosh ( 2 t) \cos ( 4 \rho t) -1}{ \cosh t \sinh (2 t)},
\end{eqnarray}
and with the lower continuum boundary 
$\omega^{L}_q
=
\frac{ \pi }{2}    \sin q$
and the upper continuum boundary
$
\omega^{U}_q
=
\pi   \sin \frac{q}{2}
$.
The auxiliary variable $\rho$ is a function of $q$ and $\omega$
\begin{eqnarray}
\rho (q, \omega)
=
\frac{4}{\pi} \mathrm{acosh} \, \sqrt{ \frac{ \omega_{U}^2 (q) - \omega^2_{L} (q)}{ \omega^2 - \omega^2_{L}(q)}} .
\end{eqnarray}
Inserting $S_2(q, \omega)$ into formula~\re{eq: master formula Al}, 
we obtain $A(r)$ in the two-spinon approximation
\begin{eqnarray} \label{A(l) two spinon}
A_{2} (r)
&=& 
  \frac{4}{ \pi^2}  
 \int_{0}^{  \pi} d q \, 
 \cos^2\frac{q}{2}   \cos ( q r )
 \\
&&  \qquad \qquad
 \times
 \int_{\omega^L_q}^{\omega^U_q }  
\frac{ d \omega'}
 { \omega^{\prime }}
 \frac{e^{-I( \rho ( q, \omega'))}}{ \sqrt{ \omega^2_{U} (q) - \omega'^2}} .
 \nonumber
\end{eqnarray}
Expression~\re{A(l) two spinon} can be evaluated numerically in 
an efficient way, provided one splits off the singular part from 
integral~\re{aux integral}
(see Ref.~\onlinecite{karbach1997}).
In doing so, we obtain for the ratio between 
the couplings $\mathcal{J}_2$ and $\mathcal{J}_1$
the following result within the two-spinon approximation
\begin{subequations}  \label{ratio}
\begin{eqnarray}  \label{ratioA}
\frac{{\mathcal J}_2}{{\mathcal J}_1}
\simeq
\frac{2 A_2(2)}{ A_2(1)}
\simeq
-0.7446,
\end{eqnarray}
and find that $\mathcal{J}_1$ is ferromagnetic,
while $\mathcal{J}_2$ is antiferromagnetic.
Similarly, the magnitudes of 
further neighbor interactions are estimated to be
\begin{eqnarray}  \label{ratioB}
\frac{{\mathcal J}_3}{{\mathcal J}_1}
& \simeq &
+ 0.1909,
\qquad
\frac{{\mathcal J}_4}{{\mathcal J}_1}
 \simeq 
- 0.2312.
\end{eqnarray}
\end{subequations}
with $\mathcal{J}_3 <0$ and $\mathcal{J}_4 > 0$.
 
The two-spinon intensity accounts for about $73\%$ of the total structure 
factor intensity.~\cite{karbach1997} The remaining part is carried by states with a higher number 
of spinons, and it is believed that the four-spinon intensity together with the two-spinon intensity cover about $98\%$ of 
the spectral weight.~\cite{caux2006} In principle, it would be possible to evaluate the 
ratios in Eq.~\re{ratio} within the four-spinon approximation. However the involved numerical integrals are rather expensive to compute. 
Instead, we shall use finite size results for the structure factor to obtain a second estimate for the
magnitude of the couplings $\mathcal{J}_r$.

\subsection{Finite size results}

We compute $A(r)$ using numerical data from the ABACUS database~\cite{cauxWeb} 
for  the dynamical structure factor in a finite system of $N=500$ sites. 
These numerical data were obtained using a method based on the Bethe ansatz framework, 
which involves a summation over the so-called determinant representations for form factors of spin operators
on the lattice.\cite{caux2005}
In these computations the momentum delta functions are smoothed out by 
including a finite broadening $\eta \sim 1/N$.
Using the numerical data with a smearing $\eta = 0.01$ we obtain for
the ratio $\mathcal{J}_2 / \mathcal{J}_1$
\begin{subequations} \label{numerics ratio}
\begin{eqnarray} 
\frac{ {\mathcal J}_2 }
{ {\mathcal J}_1  }
\simeq
\frac{ 2 A_N (2) }{ A_N (1) }
\simeq
-0.7013 ,
\end{eqnarray}
with $\mathcal{J}_1 < 0$ and $\mathcal{J}_2 > 0$, in accordance with Eq.~\re{ratioA}.
The estimated magnitudes of further neighbor interactions are  
\begin{eqnarray}
\frac{{\mathcal J}_3}{{\mathcal J}_1} 
& \simeq &
+ 0.2349,
\qquad
\frac{{\mathcal J}_4}{{\mathcal J}_1}
\simeq
- 0.1453 ,
\end{eqnarray}
\end{subequations}
where $\mathcal{J}_3$ is ferromagnetic and $\mathcal{J}_4$ is antiferromagnetic. As anticipated from the RG 
treatment in Sec.~\ref{subsec:rg}, we observe that the magnitude of the coupling strengths decreases
with spin separation distance.

Comparing Eqs.~\re{ratio} to Eqs.~\re{numerics ratio}, we infer
that the values of $\mathcal{J}_2 / \mathcal{J}_1$ agree well (within 10\%),
whereas the agreement between further neighbor interactions is worse. Since 
the two-spinon approximation misses 27\% of the structure factor intensity,
we suspect that the true values of the coupling strengths
are closer to the finite size results \re{numerics ratio} than to the two-spinon estimates.
In the next section, we are going to work out the interstitial spin ordering using
results (\ref{ratio}) and (\ref{numerics ratio}) and neglecting any longer ranged interactions 
with $\mathcal{J}_{r>4}$.

\section{Spatially anisotropic triangular lattice} \label{sec: vier}

As the discussions in the previous sections have shown, the
ordering of the dangling  spins ${\bm s}$, which occurs at  an energy scale $O[(J')^2]$,
is independent of the chain ordering.
At the energy scale $O[(J')^2]$ the spins $\bm{s}$, which  form a spatially anisotropic
triangular lattice\cite{jolicoeur90,merino99,trumper99,weihong99},
are described by the effective Hamiltonian $H_{\triangle}$, Eq.~\re{H dreieck}, with
antiferromagnetic interactions along the horizontal bonds ($J_2$) and
ferromagnetic interactions along the diagonal bonds ($J_1$), see Fig.~\ref{fig: triangular lattice}.
Neglecting third-nearest neighbor and longer ranged interactions, 
we can truncate the sums over $r$ in Eq.~\re{H dreieck} after the
first two terms and obtain
\begin{eqnarray} \label{eq: triang H}
H_{\triangle} 
&&\approx
{\mathcal J}_1  \sum_{\left[ i j \right] } \bm{s}_i \cdot \bm{s}_j
+
{\mathcal J}_2  \sum_{ \left\langle  i j \right\rangle } \bm{s}_i \cdot \bm{s}_j 
\nonumber\\
&& \qquad
+
{\mathcal J}_3  \sum_{\left[ i j \right]^{\prime} } \bm{s}_i \cdot \bm{s}_j
+
{\mathcal J}_4  \sum_{ \left\langle  i j \right\rangle^{\prime} } \bm{s}_i \cdot \bm{s}_j ,
\end{eqnarray}
where $\left[  i j \right]$ and $\left[  i j \right]'$ denote the diagonal bonds
and $\left\langle  i j \right\rangle$ and $\left\langle  i j \right\rangle'$ the horizontal bonds
connecting first and second-nearest neighbors, respectively (see Fig.~\ref{fig: triangular lattice}).
 The values of the coupling strength $\mathcal{J}_1 \cdots \mathcal{J}_4$ are
given by Eqs.~(\ref{ratio},\ref{numerics ratio}).

This is a non-trivial spin-$1/2$ quantum model.  However, it is also one
which has been heavily studied, at least in the nearest-neighbor limit.
In this case, over the vast majority of the phase diagram, the quantum ground
state agrees with the classical one.  This is particularly true when the
couplings are such that the classical ground state is ``close'' to
ferromagnetic (indeed the fully polarized ferromagnetic state is of
course an exact eigenstate, as usual).  We therefore expect that a
classical analysis is reliable, and pursue it below. 

The classical phase diagram of Hamiltonian \re{eq: triang H} is found by replacing the spin 
operators with classical coplanar spiral vectors,
\begin{eqnarray}
 \bm{s}_{i}
= \hat{x} \cos ( \bm{q} \cdot \bm{r}_i ) + \hat{y} \sin ( \bm{q} \cdot \bm{r}_i ) ,
\end{eqnarray}
 and minimizing the energy, which amounts to minimizing the Fourier transform
 of the exchange coupling\cite{villain59,yoshimori59}
 \begin{eqnarray} \label{exch coupl}
 J ( \bm{q} )
 =
 \sum_{i,j} {\mathcal J}_{i,j} \cos \left[ \bm{q} \cdot ( \bm{r}_i - \bm{r}_j ) \right] ,
 \end{eqnarray}
 where $\bm{r}_i$ denotes the position of site $i$ in real space, and
 $\hat{x}$ and $\hat{y}$ are two orthogonal unit vectors.
 In our case, Eq.~\re{exch coupl} gives
 \begin{eqnarray}  \label{exch coupl 2}
 J ( \bm{q} )
 &=&
 2 {\mathcal J}_1 \cos q_x \cos q_y + {\mathcal J}_2 \cos ( 2 q_x ) 
\\
 && \qquad
 + 2 \mathcal{J}_3 \cos ( 3 q_x ) \cos   q_y 
   + \mathcal{J}_4 \cos ( 4 q_x ).
\nonumber
 \end{eqnarray}
 with $J_1 , J_3 < 0$ and $J_2, J_4 > 0$.
 
For simplicity, let us first analyze the minima of 
Eq.~\re{exch coupl 2} for the case $\mathcal{J}_3 = \mathcal{J}_4 = 0$.
This gives $\bm{q} = (q_x , 0 )$  with
\begin{eqnarray}
q_x =
\left\{ \begin{array}{l c}
0, & \quad {\mathcal J}_2  < | {\mathcal J}_1 | /2  , \\
\arccos [ - {\mathcal J}_1  / (2  {\mathcal J}_2  ) ] , 
& \quad  {\mathcal J}_2  < | {\mathcal J}_1 | /2 .
\end{array}
\right.
\end{eqnarray}
That is, for $ {\mathcal J}_2  < | {\mathcal J}_1 | /2$  the classical ground state is ferromagnetic,  
whereas for  $ {\mathcal J}_2  > | {\mathcal J}_1 | /2$
the ground state is a spiral state rotating along the $x$ direction with the
wave vector $\bm{q} = (q_x, 0 )$.
[To study how quantum fluctuations alter the classical ground state in the case
$\mathcal{J}_3 = \mathcal{J}_4=0$ we have performed a 
linear spin-wave analysis (not shown). The magnetization decreases smoothly from 
$1/2$ to zero as $ {\mathcal J}_2 / | {\mathcal J}_1 | $ is increased, with a kink at 
$ {\mathcal J}_2 / | {\mathcal J}_1 | = 1/2$.]

In the case of non-zero further neighbor interactions, $\mathcal{J}_3
\ne 0$ and $\mathcal{J}_4 \ne 0$, there is no simple explicit expression
describing the global minima of $J ( \bm{q} )$, Eq.~\re{exch coupl
  2}. Numerically we find that the minimum of $J(\bm{q})$ with the
coupling values given by Eq.~\re{numerics ratio} (the finite size
results) occurs at $\bm{q}=0$, i.e., the ferromagnetic state is the
ground state. However, the coupling parameters of Eq.~\re{numerics
  ratio} are rather close to the boundary of a spiral phase in the
coupling parameter space $\{ \mathcal{J}_r \; | \; r=1,\cdots,4 \}$.  In
particular, the coupling parameters in Eq.~\re{ratio} (the two-spinon result)
yield a spiral state with $\bm{q} \simeq 2\pi (0.08, 0)$. Therefore, we
shall consider in what follows a cycloidal spiral ground state with
wave vector $\bm{q}=(q_x , 0)$, with $q_x$ small, which includes as a
limiting case the ferromagnetic state ($q_x =0$), that is,
\begin{eqnarray} \label{spiral2}
\hspace{0.2cm} 
&& \langle\bm{s}_{2x \pm \frac{1}{2}, 2y \mp \frac{1}{2} }\rangle
= \\
&& s_0 \left(\hat{x} \cos \left[ 2 q_x   x \pm \frac{q_x}{2}  \right]
+
\hat{y} \sin \left[ 2 q_x   x \pm \frac{q_x}{2}  \right] \right).
\nonumber
\end{eqnarray}
Here $s_0 \lesssim 1/2$ is the local static moment (staggered
magnetization) of the spiral state.

\section{Coupled Heisenberg chains in a spiral magnetic field} 
\label{sec: RG}
 
We are now in a position to address the spin-ordering pattern of the
chains.  We assume that the interstitial spins have ordered into the
spiral state given by Eq.~\re{spiral2},  and focus on the
lower temperature scale of order $(J')^4/J^3$, at which the chain spins
become affected by relevant interchain interactions.  The unperturbed
system of antiferromagnetic Heisenberg chains is given by
(note that we let $x \to x-1/2$ compared to Sec.~\ref{sec: vier})
\begin{eqnarray}
H_0
=
J \sum_{x,y} \bm{S}_{x-\frac{1}{2},y } \cdot \bm{S}_{x+\frac{1}{2} ,y}  .
\end{eqnarray}
The perturbations are: (i) coupling between chain and interchain spins,
described by \re{spiral field}, (ii) marginal ``backscattering'' term,
which is already present for a single Heisenberg chain,
and (iii) the fluctuation-generated interchain interactions 
\re{eq: interchain}.

Since the interchain spins form a two-dimensional ordered spiral state,
described in the preceding section, the main effect of their interaction
with the chain spins $\bm{S}$ is described by the spiral magnetic
field vector $\langle \bm{s}_x\rangle$ given by (note that we let $q_x
  \to q$  compared to Sec.~\ref{sec: vier})
\begin{eqnarray} \label{eq:s-spiral}
\langle \bm{s}_x\rangle
= s_0 [\hat{x} \cos (q x) + \hat{y} \sin ( q x )] ,
\end{eqnarray} 
with $q$ small.
The spiral magnetic field introduces the following perturbations to
this system of decoupled spins 
\begin{eqnarray}  \label{spiral field}
H_S
&=&
J'
\sum_{x,y}   \Big\{ 
\bm{S}_{2x-1/2,y} \cdot \langle \bm{s}_{2x-1} + \bm{s}_{2 x} \rangle
\\
&& \qquad \qquad \qquad
+
\bm{S}_{2x+1/2,y} \cdot \langle \bm{s}_{2x} + \bm{s}_{2 x+1} \rangle
\Big\} 
\nonumber \\
& = & \sum_{x,y} {\bf h}_{x+1/2} \cdot {\bf S}_{x+1/2,y},
\end{eqnarray}
with
\begin{equation}
  \label{eq:19}
{\bf h}_x = 2 h_0 \cos(q/2) [\hat{\bf x} \cos (q x) + \hat{\bf y} \sin ( q x )] ,
\end{equation}
and the field  strength $h_0 = s_0 J' $.

The simplest effect of the spiral field, as for any field, is to induce
a corresponding spiral magnetization in linear response.  This implies
that, generically, there will be static components of the chain spins in
the spiral ($x$-$y$) plane.  Since the spiral field is proportional to
$J'$, and the chain susceptibility is generally proportional to $1/J$,
these static components are of order $J'/J$.  

This simple linear response, however, is not the complete story.  A much
more detailed analysis is needed to resolve the more subtle effects of
the spiral field beyond linear response, in conjunction with the
inter-chain couplings described in Sec.~\ref{sec: zwei}.  The nature of
this more complete analysis depends crucially upon the
magnitude of $q$.  When $q$ is small, the spins respond in a way
which is similar to the response to a uniform magnetic field.  A
systematic approach is then possible, in which the Hamiltonian is
transformed into the slowly rotating frame in which the external field
is uniform.  This is tractable because for small $q$ the
non-Heisenberg interactions induced by the change of frame are weak.
Because of the expected smallness of $q$, we focus on this case in the
remainder of this section.

In the opposite limit of a ``large'' wave vector, which is
incommensurate with the dominant fluctuations of the 1D Heisenberg
chain, i.e. $q,|\pi-q| \sim O(1)$, the field weakly couples to the
spin chain.  Indeed, the leading $O(h)$ effects average out over space, and
instead only sub-dominant terms are induced at $O(h^2)$.  In this limit
these weaker $O(h^2)$ terms are crucial in determining the final state
of the system.  Though this limit is actually conceptually simpler than
the opposite one, it is technically challenging, because the effects of
spiral field are determined by very short-wavelength properties of the
Heisenberg spins.  As a result, we are not unambiguously able to resolve
the ground state in this limit.  However, the ambiguity is small: we
show in Appendix~\ref{sec:large-q} that the system at zero temperature
is in  one of only {\sl two} possible phases.  One of these is the {\sl
  same} non-coplanar state which we find in the small $q$ limit (the
other is a coplanar state).  This supports the notion that, at least up
to some critical $O(1)$ value of $q$ (and possibly for all $q$), the
ground state evolves smoothly from the small $q$ limit.

\subsection{Transformation to rotated frame}
\label{sec:rotated frame}

From here on, we assume $q\ll 1$.  It is advantageous to rotate the
chain spins towards the direction of the spiral magnetic field
\begin{eqnarray} \label{first rot}
&&
\bm{S}_{x,y}
\longrightarrow
\mathcal{R}_x 
\bm{S}_{x,y},
\nonumber\\
&&
\mathcal{R}_x
=
\begin{pmatrix}
0 & + \sin ( q x) & - \cos (  q x) \cr
0 & -\cos (  q x) & - \sin (  q x) \cr
-1 & 0 & 0 \cr
\end{pmatrix} .
\end{eqnarray}
The rotation $\mathcal{R}_x$ amounts to $-\pi/2$ rotation about the
$y$ axis followed by a rotation about the $z$ axis with angle $qx$.  We
find that under rotation \re{first rot} the magnetic field term becomes
\begin{eqnarray}
\tilde{H}_S
=
- 2 h_0 \cos ( q /2 ) \sum_{x,y}  S^z_{x + 1/2,y} .
\end{eqnarray}
$H_0$ transforms into  
\begin{eqnarray} \label{transformed H0}
&&
H_0 = J \sum_{x,y} \Big( S^x_{x-\frac{1}{2},y } S^x_{x+\frac{1}{2},y }
+ \cos q \big[ S^y_{x-1/2,y} S^y_{x+1/2,y} 
\nonumber\\
&& \quad  
+ S^z_{x-1/2,y} S^z_{x+1/2,y} \big]
+  \sin q  \big[ S^y_{x-1/2,y} S^z_{x+1/2,y} 
\nonumber\\
&& \quad  
- S^z_{x-1/2,y} S^y_{x+1/2,y} \big] \Big).
\end{eqnarray}
In the limit of small $q$, which we focus on,
this Hamiltonian is conveniently split into  Heisenberg one $\tilde{H}_0$ with
the modified exchange constant $\tilde{J} = J \cos q$, 
\be
\tilde{H}_0 = \tilde{J} \sum_{x,y} \bm{S}_{x-\frac{1}{2},y } \cdot \bm{S}_{x+\frac{1}{2},y } ,
\ee
an effective Ising anisotropy
$\tilde{H}_1$ along the $x$ axis,
\be \label{eq:H1}
\tilde{H}_1 = J (1 - \cos q) \sum_{x,y} S^x_{x-\frac{1}{2},y } S^x_{x+\frac{1}{2},y } ,
\ee
and an effective Dzyaloshinskii-Moriya (DM) interaction $\tilde{H}_2$
\be \label{eq:H2}
\tilde{H}_2 = J \sin q \sum_{x,y} 
\left[
S^y_{x-1/2,y} S^z_{x+1/2,y} 
- S^z_{x-1/2,y} S^y_{x+1/2,y} .
\right]
\ee
Thus, $H_0 = \tilde{H}_0 + \tilde{H}_1 + \tilde{H}_2$.

\subsubsection{Low-energy limit}
\label{sec:low-energy-limit}

It is appropriate now to take a low-energy limit, for which we
use the non-Abelian spin current formulation.  The zeroth-order
Heisenberg Hamiltonian in the continuum limit yields the fixed point
term plus a backscattering correction, $\tilde{H}_0 \rightarrow \tilde{H}_0 +
H_{bs}$.   The fixed point term, written in the Sugarawa form, is
\begin{equation}
  \label{eq:26}
  \tilde{H}_0 = \frac{ 2 \pi \tilde{u}}{3} \sum_y \int dx
  \left[ \bm{J}_{L,y} \cdot \bm{J}_{L,y} + \bm{J}_{R,y} \cdot \bm{J}_{R,y} \right] ,
\end{equation}
where $\tilde{u} = \cos q u$ is the modified spinon velocity.  To the
order we work in this section, it is sufficient to take
$\tilde{u}\approx u = \pi J/2$.  The backscattering correction is
\begin{eqnarray}
H_{bs}
&=&
\sum_y
\int d x ~\tilde{g}_{bs} \bm{J}^{\ }_{L,y} \cdot \bm{J}^{\ }_{R,y} ,
\label{eq:bs}
\end{eqnarray}
where $\tilde{g}_{bs} = \cos q g_{bs}< 0$.  Again we can approximate
$\tilde{g}_{bs} \approx g_{bs}$ here.    

The Ising anisotropy can also be expressed in terms of currents.  One
must take care since it is a composite operator.  One obtains 
\begin{eqnarray}
&&
\tilde{H}_1 = (1-\cos q) \sum_y \int dx \Big[ \frac{2\pi u}{3}
(J^x_{L,y} J^x_{L,y} + J^x_{R,y} J^x_{R,y} ) 
\nonumber\\
&& \qquad
+ g_{bs} J_{R,y}^x J_{L,y}^x  
\Big] .
\label{eq:delta1}
\end{eqnarray}
The DM  (see Ref.~\onlinecite{gangadharaiah2007} for details)
and external field $\tilde{H}_S$ terms add up to 
\begin{equation} \label{eq 520}
\tilde{H}_2 + \tilde{H}_S  
= 
\sum_y \int dx
\left[  \tilde{d}( J^x_{R,y} - J^x_{L,y}) - 
\tilde{h} (J^z_{R,y} + J^z_{L,y})
\right] ,
\end{equation}
where $\tilde{d} = ( 3 / \pi ) \sin q J$
and $\tilde{h} = 4 h_0 \cos (q/2)$.  
We will consider both  
contributions in Eq.~\re{eq 520} on equal footing. 
 Formally, we consider $q\ll 1$ and
$J'/J \ll 1$, but with $q J/J'$ arbitrary.  In this limit we may
approximate $\tilde{J} \simeq J$,  $\tilde{d} \simeq 3 q J/\pi$ and $\tilde{h} \simeq 2h_0$.
Moreover, $\tilde{H}_1$ can be dropped completely, since it represents
next, $q^2$, order anisotropy corrections to both $H_0$
and the marginal backscattering term $H_{bs}$.
(In principle,  terms of order $q^2$ could
be included by considering velocity shifts and small anisotropy corrections to the
backscattering coupling $g_{bs}$. However, when $q$ is small enough, 
these higher-order corrections will not affect the outcome of the analysis in an essential way.)

Finally, the relevant interchain interactions read
\begin{eqnarray} \label{eq:H-2prime}
H''
&=&
\sum_y
\int dx 
\left\{
\gamma_N \bm{N}_y \cdot \bm{N}_{y+1}
+
\gamma_{\varepsilon} \varepsilon_y \varepsilon_{y+1}
\right\} .
\end{eqnarray}
The analysis in Sec.~\ref{sec: zwei} shows that coupling
constants $\gamma_{N,\varepsilon}$ are of order $(J')^4/J^3$, and
importantly, $\gamma_N>0$. 

\subsection{Chiral SU(2) rotation}

An unique feature of the WZNW field theory is its emergent {\sl chiral}
symmetry under independent SU(2) rotations for the left and right moving
sectors.  We take advantage of this to remove the DM term in Eq.~\re{eq
  520}.  Specifically, we rotate the right and left chiral spin currents
about the $y$ axis by {\sl opposite} angles, $+\theta$ and
$-\theta$, respectively.
\begin{eqnarray} \label{RL rot 2}
\bm{J}_{R,y} 
\longrightarrow \mathcal{R}_R \bm{J}_{R,y},
\qquad
\bm{J}_{L,y} \longrightarrow \mathcal{R}_L \bm{J}_{L,y},
\end{eqnarray}
with
\begin{eqnarray} 
\mathcal{R}_{R/L}
&=&
\begin{pmatrix}
\cos (\theta)  & 0 & \mp\sin (\theta)  \cr
0 & 1 & 0 \cr
\pm\sin (\theta)  & 0 & \cos (\theta)  \cr
\end{pmatrix} ,
\end{eqnarray}
where $\theta = + \atan ( \tilde{d} / \tilde{h} )$.
Under the rotations \re{RL rot 2}, the staggered magnetization and dimerization transform to
\begin{eqnarray}
N^{x,z}_y
&  \rightarrow &
N^{x,z}_y,
\nonumber\\
N^y_y
&  \rightarrow &
\cos \theta \, N^y_y  + \sin \theta \, \varepsilon_y,
\\ 
\varepsilon_y
& \rightarrow &
\cos \theta \, \varepsilon_y - \sin \theta \, N^y_y .
\nonumber
\end{eqnarray}
Due to the chiral SU(2)$_R \times$ SU(2)$_L$ symmetry ,  the low-energy
$\tilde{H}_0$ Hamiltonian is unaffected. 
Vector perturbation \re{eq 520} simplifies to $\tilde{H}_z$, where
\be
\tilde{H}_z = - \sqrt{ ( \tilde{h} )^2 + ( \tilde{d} )^2 } \sum_{y} \int dx ( J^z_{L,y} + J^z_{R,y} )  .
\label{eq:H_z}
\ee
Under the rotation the back-scattering term transforms into
\begin{eqnarray} \label{eq:rotated bs}
&&
H_{bs}
=
\sum_{x,y} \Big\{
\frac{g_1}{2} ( J^+_{L,y} J^{-}_{R,y} + J^-_{L,y} J^+_{R,y} )
\\
&& \quad
+
\frac{g_2}{2} ( J^+_{L,y} J^+_{R,y} + J^{-}_{L,y} J^{-}_{R,y} )
+
g_4 J^z_{L,y} J^z_{R,y} 
\nonumber\\
&& \quad
+
\frac{ g_3}{2} ( J^z_{L,y} J^+_{R,y} + J^z_{L,y} J^-_{R,y} - J^{+}_{L,y} J^z_{R,y}
- J^{-}_{L,y} J^z_{R,y} )
\Big\} ,
\nonumber
\end{eqnarray}
where the couplings $g_i$ can be expressed in terms of the backscattering $g_{bs}$
\begin{eqnarray} \label{g2 and g4}
&&
g_1
=
\frac{g_{bs} }{2} ( 1 + \cos 2 \theta ) ,
\qquad
g_2
=
\frac{g_{bs} }{2} ( \cos 2 \theta - 1 ),
\nonumber\\
\\
&&
g_3 = g_{bs} \sin 2 \theta ,
\qquad \qquad \quad
g_4 =  g_{bs} \cos 2 \theta .
\nonumber
\end{eqnarray}
The interchain perturbation $H''$ is significantly affected as well, and now reads
\begin{eqnarray} \label{eq: H prime prime}
H'' 
&=&
\sum_{x,y} 
\Big\{
\cos^2 \theta ( 
\gamma_{\varepsilon} \varepsilon_y \varepsilon_{y+1}
+ \gamma_{N} N^y_y N^y_{y+1}
)
\nonumber\\
&& \qquad    
+ \sin^2 \theta (
\gamma_{N} \varepsilon_y \varepsilon_{y+1}
+ \gamma_{\varepsilon} N^y_y N^y_{y+1}
)
\\
&& \qquad    +
(\gamma_{N} - \gamma_{\varepsilon} ) \cos \theta \sin \theta
( \varepsilon_{y} N^y_{y+1} + N^y_y  \varepsilon_{y+1} )
\nonumber\\
&& \qquad 
+
\gamma_{N} N^x_y N^x_{y+1}
+
\gamma_{N} N^z_y N^z_{y+1}
\Big\} 
.
\nonumber
\end{eqnarray}

\subsection{Absorption of the field $h^z$}
\label{sec:absorption-field-hz}

The benefit of the chiral rotation is that 
within Abelian bosonization\cite{shankar,nagaosa,senechal1999} we can now absorb the 
magnetic field $h^z : =  \sqrt{ ( \tilde{h} )^2 + ( \tilde{d} )^2 }$ by the usual shift
\cite{affleck99,bogoliubov86}
\begin{eqnarray} \label{eq: field shift}
\varphi_{s,y}
\rightarrow 
\varphi_{s,y} +  \frac{1}{ \sqrt{ 2 \pi } } \frac{h^z}{u}  x ,
\end{eqnarray}
for all $y$.  Note that $h^z = J'\sqrt{(2s_0)^2 + (3 q J/\pi J')^2}$ is
$O(J')$ in the scaling limit considered here [see the discussion after
Eq.~\re{eq:bs}].  

This transforms the currents in the following way
\begin{eqnarray}
J^{\pm}_{L,y} & \rightarrow & J^{\pm}_{L,y} e^{ \pm i \frac{h^z}{u}  x} ,
\qquad
J^{\pm}_{R,y} \rightarrow  J^{\pm}_{R,y} e^{ \mp i \frac{h^z}{u}  x} ,
\\
J^z_{L,y} & \rightarrow & J^z_{L,y} + \frac{h^z}{4\pi u}, 
\qquad J^z_{R,y}  \rightarrow J^z_{R,y} + \frac{h^z}{4\pi u}.
\end{eqnarray}
The latter transformation explicitly embodies the linear response of the
chain magnetization $M_y^z = J_{R,y}^z+J_{L,y}^z$ to the field.

The scaling
dimension $1/2$ fields transform according to
\begin{eqnarray}
  \label{eq:12}
  N^z_{y} & \rightarrow & \cos \left( \frac{h^z}{u} x \right) N_y^z 
  +  \sin \left( \frac{h^z}{u} x \right) \epsilon_y, \\
  \epsilon_{y} & \rightarrow & \cos \left( \frac{h^z}{u} x \right) \epsilon_y 
  -  \sin \left( \frac{h^z}{u} x \right) N^z_y, 
\end{eqnarray}
and $N_y^x$ and $N_y^y$ remain unchanged.  Making this shift renders several
terms in Eqs.~(\ref{eq:rotated bs},\ref{eq: H prime prime}) oscillatory,
at scales $x> u/h^z$.

\subsection{Renormalization group equations}

Now we consider the effect of the various couplings.  Because of the
explicit oscillatory factors introduced by the shift in Eq.~(\ref{eq:
  field shift}), we must consider two separate regimes of the flow.
First, on scales shorter than the period of these oscillations, the
oscillations themselves can be neglected, and we should consider all the
couplings in Eqs.~(\ref{eq:rotated bs},\ref{eq: H prime prime}).  On
longer scales, the oscillatory couplings may be dropped entirely.  The
reader may be familiar with a similar treatment of the effects of a
field on a Heisenberg chain by Affleck and Oshikawa.\cite{affleck99}

\subsubsection{Short-scale flows}
\label{sec:short-scale-flows}

Consider first the short-scale flows, i.e., the regime when
$\frac{h^z}{u} e^\ell a_0 \lesssim 1$.  This means
\begin{equation}
  \label{eq:17}
   0\leq \ell \lesssim \ell^* =  \ln
\frac{u}{h^z a_0} \sim \ln (J/J').
\end{equation}
We neglect completely the effect of
the oscillatory factors induced in $H_{bs}$ and $H''$.    Although the form in
Eq.~(\ref{eq:rotated bs}) is complicated, the flows remain simple.  This
is because Eqs.~(\ref{eq:rotated bs},\ref{g2 and g4}) are obtained from
the chiral SU(2) rotation which is a symmetry of the fixed point
Hamiltonian.  Thus, since the field $h^z$ has no effect in this energy
range, the flows remain fully SU(2) symmetric.  They are simply
\begin{eqnarray}
  \label{eq:14}
  \frac{dg_{bs}}{d\ell} & = & \frac{[g_{bs}(\ell)]^2}{2\pi u},
\end{eqnarray}
and
\begin{eqnarray}
  \label{eq:16}
  \frac{d\gamma_N}{d\ell} & = & \gamma_N - \frac{1}{4\pi u}g_{bs} \gamma_N,
  \\
  \frac{d\gamma_\varepsilon}{d\ell} & = & \gamma_\varepsilon
  +\frac{3}{4\pi u}g_{bs} \gamma_\varepsilon .
\end{eqnarray}
One can check this simple result by directly calculating the flow
equations for $g_1\cdots g_4$, and showing that the forms in
Eqs.~(\ref{g2 and g4}) are preserved by these equations.  This is the
result of the simple argument above.

We note that it is sufficient to work only to linear order in the
relevant couplings $\gamma_N,\gamma_\varepsilon$, since their initial
values are of order $(J')^4$, and therefore remain small over this range
of $\ell$ (they increase only by a factor of $e^\ell \sim J/J'$).
To this order, they do not feed back into the flow of $g_{bs}$.  The
usual solution to Eq.~(\ref{eq:14}) obtains:
\begin{eqnarray}
  \label{eq:15}
  g_{bs}(\ell) & = & 
  \frac{g_{bs}}{(1-\frac{g_{bs}}{2\pi u}\ell)} . 
\end{eqnarray}
Since $g_{bs}<0$, it becomes small under renormalization, and
specifically of order $u/\ell$ for $\ell \gg 1$.
Inserting this into the remaining equations and solving gives
\begin{subequations}
\begin{eqnarray}
  \label{eq:13}
  \gamma_N(\ell) & = & \left(1- \frac{g_{bs} \ell}{2\pi
      u}\right)^{1/2} e^\ell \,\gamma_{N}(0) , \\
  \gamma_\varepsilon(\ell) & = &  \left(1- \frac{g_{bs} \ell}{2\pi
      u}\right)^{-3/2} e^\ell \,\gamma_{\varepsilon}(0) .
\end{eqnarray}
\end{subequations}
Evaluating this at $\ell=\ell^*$ and using Eqs.~(\ref{ap:relevant}) for
the initial conditions gives
\begin{subequations}   \label{eq:18}
\begin{eqnarray}
  \gamma_N(\ell^*) & \sim & \frac{1}{2\pi^4}\left(\frac{|g_{bs}| \ln (J/J')}{2\pi
      u}\right)^{1/2}  \frac{(J')^3}{J^2} , 
      \\
  \gamma_\varepsilon(\ell^*) 
  & \sim & 
  -\frac{3}{4 \pi^4}\left(\frac{|g_{bs}| \ln (J/J')}{2\pi
      u}\right)^{-3/2}  \frac{ (J')^3}{ J^2} .
\end{eqnarray}
\end{subequations}
Note that these couplings indeed remain small at this scale.
Furthermore, the staggered magnetization coupling $\gamma_N$ is already
parametrically enhanced over the dimerization coupling
$\gamma_\varepsilon$, by a factor of $\ln^2 (J/J')$.  

\subsubsection{Long-scale flows}
\label{sec:long-scale-flows}

Now we consider the renormalization on scales longer than the period of
the oscillations induced by the field shift.  Here the SU(2) symmetry
is truly broken, and the RG deviates from the simple one above.  We drop
all oscillating terms (this includes $g_1$, $g_3$ and several of the terms
in $H''$), so that the remaining perturbations to the bare Hamiltonian
$\tilde{H}_0$ are
\begin{eqnarray} \label{non-oscill perturb}
H_{bs}
&=&
\sum_{x,y} 
\left\{
\frac{\tilde{g}_2}{2} ( J^+_{L,y} J^{+}_{R,y} + J^{-}_{L,y} J^{-}_{R,y} )
+
\tilde{g}_4 J^z_{L,y} J^z_{R,y}
\right\} ,
\nonumber\\
H''
&=&
\sum_{x,y}
\Big\{
\tilde{\gamma}_{N^y} N^y_y N^y_{y+1} + \tilde{\gamma}_{N^x} N^x_y N^x_{y+1} 
\\ 
&& \qquad  \qquad
+
\tilde{\gamma}_+ ( N^z_y N^z_{y+1} + \varepsilon_y \varepsilon_{y+1} )
\Big\} ,
\nonumber
\end{eqnarray}
where we have introduced the new coupling constants
$\tilde{g}_2,\tilde{g}_4,\tilde{\gamma}_{N^x}, \tilde{\gamma}_{N^y},
\tilde{\gamma}_+$.  They should be matched at $\ell=\ell^*$ to the
couplings from the short-scale flows, defined in Eqs.~(\ref{eq: H prime
  prime},\ref{eq:rotated bs}), which implies 
\begin{eqnarray} \label{def tilde couplings}
\tilde{g}_2(\ell^*) &=& - g_{\text{bs}}(\ell^*)  \sin^2 \theta , \nonumber\\
\tilde{g}_4(\ell^*)  &=& g_{\text{bs}}(\ell^*)  \cos 2\theta ,\nonumber\\
\tilde{\gamma}_{N^x}(\ell^*) 
&=&
\gamma_{N}(\ell^*)  , 
\nonumber\\
\tilde{\gamma}_{N^y}(\ell^*)  & = &
\cos^2 \theta \gamma_{N}(\ell^*) 
+
\sin^2 \theta \gamma_{\varepsilon}(\ell^*) ,
\\
\tilde{\gamma}_+(\ell^*) 
&=&
\left[
\cos^2 \theta \,\gamma_{\varepsilon}(\ell^*) 
+ 
(1+\sin^2 \theta) \gamma_{N}(\ell^*) 
\right]/2 .
\nonumber
\end{eqnarray}
We now compute the RG equations for the perturbations 
\re{non-oscill  perturb} to the bare Hamiltonian $H_0$ using well-established
OPEs for the non-Abelian spin currents (see for example
Ref.~\onlinecite{starykh2005}).  After some lengthy calculations we find
\begin{eqnarray} \label{eq:final-rg}
\frac{d\tilde{g}_2}{d\ell} & = & - \frac{\tilde{g}_4\tilde{g}_2}{2\pi
  u}, \qquad \frac{d\tilde{g}_4}{d\ell}  =  - \frac{\tilde{g}^2_2}{2\pi
  u}, \nonumber \\
\frac{ d \tilde{\gamma}_{N^x} }
{d \ell}
&=&
\left( 1 -  \frac{1}{4 \pi u} \tilde{g}_4 +  \frac{1}{2 \pi u } \tilde{g}_2  \right) \tilde{\gamma}_{N^x},
\nonumber\\
\frac{ d \tilde{\gamma}_{N^y} }{ d \ell}
&=&
\left( 1 -  \frac{1}{4 \pi u} \tilde{g}_4 - \frac{1}{2 \pi u} \tilde{g}_2  \right) \tilde{\gamma}_{N^y} ,
\\
\frac{ d \tilde{\gamma}_+ }{ d \ell}
&=&
\left( 1 + \frac{1}{4 \pi u}  \tilde{g}_4 \right) \tilde{\gamma}_+ .
\nonumber
\label{eq:final-rg}
\end{eqnarray}
It is important to understand how these equations lead to an
instability.  The equations for $\tilde{g}_2,\tilde{g}_4$ are decoupled,
and can be solved separately.  They have the standard form found in the
Kosterlitz-Thouless (KT) analysis.\cite{gogolin1998}  One recalls that the quantity
\begin{equation}
  \label{eq:20}
  Y = \tilde{g}_2^2-\tilde{g}_4^2
\end{equation}
is a constant of the motion.  The flows are {\sl unstable} provided
$\tilde{g}_4(\ell^*) < |\tilde{g}_2(\ell^*)|$, which is {\sl always}
satisfied except when $\theta=\pi/2$ exactly, at which point it becomes
an equality.  That is for $\theta \in [0, \pi/2 [$, the trajectories tend to 
$\tilde{g}_4 \to - \infty$ and $\tilde{g}_2 \to + \infty$.
We fix $Y$ by the initial conditions,
\begin{equation}
  \label{eq:21}
  Y = |g_{bs}(\ell^*)|^2 \left(\sin^4\theta - \cos^2 2\theta\right).
\end{equation}
Hence,  $Y$ is negative for $\theta \in [0, \acos \sqrt{2/3} ]$   and 
positive for $\theta \in [ \acos \sqrt{2/3} , \pi /2 ]$.
Writing $\tilde{g}_2^2=Y+\tilde{g}_4^2$ we can solve the KT equations for
$\tilde{g}_4$.  When  $Y > 0$, i.e., in the crossover regime of the KT flow, we have
\begin{eqnarray}
  \label{eq:22}
  \atan \left(\frac{\tilde{g}_4(\ell)}{\sqrt{Y}}\right) & = & \atan
  \left(\frac{g_{bs}(\ell^*)\cos 2\theta}{\sqrt{Y}}\right) -
  \frac{\sqrt{Y}}{2\pi u} (\ell-\ell^*).\nonumber \\
\end{eqnarray}
The coupling $\tilde{g}_4$ clearly diverges when the right-hand side of
this equation reaches $\pi/2$ plus an integer times $\pi$.  The ``time'' $\ell_d$ of
this divergence is
\begin{eqnarray}
  \label{eq:24}
  \ell_d & = & \ell^* + \frac{2\pi u}{|g_{bs}(\ell^*)|} \frac{\pi/2-\atan
  \left(\frac{\cos 2\theta}{\Upsilon}\right)}{\Upsilon},
\end{eqnarray}
where we define $\Upsilon=\sqrt{Y}/|g_{bs}(\ell^*)| $.
Using  $\ell^* \sim \ln(J/J')$ and $g_{bs}(\ell^*) \sim 2\pi
u/\ell^*$  we obtain
\begin{eqnarray}
  \label{eq:25}
  \ell_d& = & \ln (J/J') \left[1 + \frac{\pi/2-\atan
  \left(\frac{\cos 2\theta}{\Upsilon}\right)}{\Upsilon}\right].
\end{eqnarray}
One can check that $\ell_d/\ln(J/J')$ increases monotonically from $4$
when $\theta= {\rm acos}\sqrt{2/3}$   to infinity as $\theta
\rightarrow \pi/2$. Similarly, when $Y < 0$ (strong-coupling regime)
  $\tilde{g}_4 (\ell) $ diverges at the length scale
\begin{eqnarray}
\ell_d 
=
\ln ( J / J' )
\left[ 1 + \atanh \left( \frac{  \widebar{\Upsilon} }{  \cos 2 \theta } \right) ( 1 / \widebar{\Upsilon} ) \right],
\end{eqnarray}
where $\widebar{\Upsilon}=\sqrt{-Y}/|g_{bs}(\ell^*)| $. In this regime $\ell_d / \ln ( J / J' )$
takes the value $4$ at $\theta = \acos \sqrt{ 2 / 3}$, increases monotonically 
with decreasing $\theta$, and diverges at $\theta = 0$.

With the values at $\ell=\ell^*$ given by Eqs.~(\ref{eq:18}, \ref{def
  tilde couplings}) the relevant couplings $\tilde{ \gamma }^{\
}_{N^x}$, $\tilde{ \gamma }^{\ }_{N^y}$, and $\tilde{\gamma}_+$ become
of order 1 at the length scale $\ell_o \sim 4 \ln ( J / J')$
[so that $\ell_o-\ell^*\sim 3\ln(J/J')$], 
which is always smaller than the scale
$\ell_d$.  The most relevant of these turns out to be
$\tilde{\gamma}_{N^x}$, as can be seen, e.g., by examining the ratio
\begin{eqnarray}
\frac{ \tilde{\gamma}_{N^x}(\ell) }{ \tilde{\gamma}_{N^y}(\ell)}
=
\frac{ \tilde{\gamma}_{N^x}(\ell^*)  }{ \tilde{\gamma}_{N^y}(\ell^*) }
\exp \left[
+ \frac{1}{\pi u} \int_{\ell^*}^\ell d x \, \tilde{g}_2(x) 
\right]  ,
\end{eqnarray}
where (remember that $g_{bs} < 0$)
\begin{eqnarray}
\tilde{g}_2(\ell^*) &=& |g_{\text{bs}}(\ell^*)| \sin^2 \theta=\frac{
  |g_{\text{bs}}(\ell^*)| \tilde{d}^2}{\tilde{d}^2 + \tilde{h}^2} . 
\label{eq:g2}
\end{eqnarray} 
From this clearly $\tilde{\gamma}_{N^x}$ becomes parametrically larger
than $\tilde{\gamma}_{N^y}$ under renormalization (for $q\neq 0$).
Similar analysis shows that $\tilde{\gamma}_{N^x}$ is also enhanced
relative to $\tilde{\gamma}_+$.  
Thus, for $q > 0$, we find that the staggered magnetization
$\tilde{\gamma}_{N^x} = \gamma_{N^x} $ dominates. Hence, the spins align
antiferromagnetically along the $x$ direction,
$\hat{\bm{S}}^{\mathrm{x}}_x$, in the rotated (``comoving'') coordinate
frame. 

\subsubsection{Ordering pattern of chain spins}
\label{sec:order-chain}

Let us now infer what this means in terms of the original spins,
in the fixed coordinate frame.  It is necessary to trace back the
transformations of the spin operators in Eqs.~(\ref{first rot},\ref{RL
  rot 2},\ref{eq: field shift}).  This is straightforward but tedious.
We will not give a general expression of the relation of the microscopic
spins to the continuum operators after the final transformation, which
is not illuminating.  Instead, we give the result for the {\sl
  expectation value} of the spin operators given that, as argued above,
in the rotated variables the ordering is very simple:
\begin{eqnarray}
  \label{eq:10}
  \langle N_{y}^a \rangle & = & M (-1)^{y} \delta^{a,x}, \\
  \langle J_R^a\rangle & = & \langle J_L^a \rangle =\langle
  \epsilon\rangle = 0.
\end{eqnarray}
Here $M\neq 0$ represents the spontaneous moment, and will become the
staggered magnetization.  The $(-1)^y$ factor obtains because
$\tilde{\gamma}_{N^x}>0$.

Now we relate the spin operators
as described above to the continuum fields:
\begin{eqnarray}
  \label{eq:11}
  \langle S_{x+\frac{1}{2},y}^x \rangle & = & 
  - \frac{h^z}{2\pi u}\cos
  \theta  \cos q x , \\
  \langle S_{x+\frac{1}{2},y}^y \rangle & = &
 - \frac{h^z}{2\pi u}\cos
  \theta  \sin q x , \\
  \langle S_{x+\frac{1}{2},y}^z \rangle & = & -(-1)^{x+y} M.
\end{eqnarray}
Thus, indeed the dominating $\gamma_{N^x}$-term orders the chain spins
antiferromagnetically along the $z$ axis perpendicular to the \mbox{$x$-$y$}
$\bm{s}$-spiral plane.  The non-zero components of the spins within the
$x$-$y$ plane are induced by the local field arising from the ordered
interstitial moments.  Note that they are of $O(J'/J)$  {\sl which is
  much larger than $M\sim (J'/J)^2$}.  Thus the static
moments on the chains are predominantly in the the plane of the spiral,
with a smaller staggered component in the perpendicular ($z$)
direction (see Fig.~\ref{fig: kagome order}).  

The case of \emph{ferromagnetic} order amongst the interchain spins
is obtained by setting $q=0$. This implies $\tilde{d} =0$ so that
$\tilde{g}_2=0$, see \re{eq:g2}.  The chains are subjected to the
uniform magnetic field $\tilde{h}$ only. This leads to $\theta=0$ and,
as a result, symmetry in the $N^x - N^y$ plane: $\tilde{\gamma}_{N^x} /
\tilde{\gamma}_{N^y} =1$.  The chain spins ${\bm S}$ order
nearly collinearly with a ferromagnetic
component of order $O(J'/J)$ and smaller antiferromagnetic component
of order $O[ ( J'/J)^2 ]$ perpendicular to the ferromagnetic moments.
 Note that in this case, because the
interstitial spins and the predominant moment of the chain spins
are ferromagnetically ordered and hence collinear,
the full magnetic order is actually {\sl coplanar}.  For instance, if
the ferromagnetic moments are aligned in the $\hat{\bf x}$ direction, the
antiferromagnetic component of the chain spins will be aligned along some axis $\hat{\bf
  e}$ in the $y$-$z$ plane, and all the spins are contained in the plane
spanned by $\hat{\bf x}$ and $\hat{\bf e}$.
Furthermore, 
 this state is ferrimagnetic, i.e., has a macroscopic net moment,
since the moments of the ferromagnetically ordered interstitial spins
 are unequal to the opposing in-plane component of the chain spin
 moments.

\section{Summary and Discussion} \label{sec: conclusions}

In this work we have analyzed the ground state phase of the quantum
Heisenberg antiferromagnet on the kagome lattice with spatially
anisotropic exchange. We have studied this problem in the quasi-1D
limit, where the lattice is broken up into antiferromagnetic spin-1/2
chains that are weakly interacting via intermediate ``dangling'' spins
$\bm{s}$ (see Fig.~\ref{fig: kagome lattice}). This limit lends itself
to a perturbative RG analysis in the weak exchange interaction $J'$
using bosonization and current algebra techniques. We find that there is
a natural separation of energy scales: the intermediate spins order at
an energy scale of order $(J')^2/J $, at which the chains are not
influenced by the interactions among themselves and with the
interstitial spins. The low-energy behavior of the chains, on the other
hand, is only modified at $O[(J')^4]$, as geometric frustration prevents
the generation of relevant interchain interactions at the larger scale
$(J')^2/J$.  We have used perturbative and numerical methods to
determine the effective interactions $\mathcal{J}_i$ among the
interstitial spins $\bm{s}$, which arise at $O[(J')^2]$. It turns out
that the spins $\bm{s}$ order in a coplanar cycloidal spiral with
wave vector $\bm{q}$ parallel to the chain direction.  The magnitude of
the wave vector is presumably rather small (if not vanishing). It depends
sensitively on the strength of further neighbor interactions among the
interstitial spins, which cannot be reliably determined from our approach.
 The ordered interstitial moments induce a spiral order of the chain spins of $O(J'/J)$.
Besides this, the chain spins exhibit a small antiferromagnetic component
of $O[(J'/J)^2]$ that points along the axis perpendicular to the spiral plane.
This \textit{non-coplanar} ground state of the spatially anisotropic kagome antiferromagnet is
illustrated in Fig.~\ref{fig: kagome order}.

It is interesting to compare to recent results for this lattice in the
spatially anisotropic limit.  Wang
\textit{et al.}\cite{wang2007} found a coplanar ferrimagnetic
chirality stripe order using a semiclassical analysis (see also
Ref.~\onlinecite{yavorskii2007}).  In this state, the interstitial spins
are ferromagnetically ordered, and the chain spins are ordered 
in an antiferromagnetic fashion, nearly collinearly along an axis 
perpendicular to the interstitial moments, but canted slightly in that direction.  
The ordering of the interstitial spins is very close to our findings; i.e.,
it corresponds to the special case $q=0$, which as we have
described cannot be excluded by our calculations.
However, even in this case the ordering pattern of the chain spins 
is quite different from  that in Ref.~\onlinecite{wang2007}, insofar as we find 
that the chain spins have a predominant \emph{ferromagnetic} component 
antiparallel to the interstitial spins (and only a considerably smaller antiferromagnetic 
component perpendicular to the interstitial spins),  
while Wang \textit{et al.}\cite{wang2007} obtained a predominantly 
\emph{antiferromagnetic} ordering among the chain spins.
 
Yavors'kii \emph{et al.} in Ref.~\onlinecite{yavorskii2007}, on the
other hand, used a large-$N$ expansion applied to the Sp($N$)-symmetric
generalization of the model.  In the limit $J' \ll J$ they found that the
chains are completely decoupled, and the interstitial spins show
some (short range) spin-spin correlation that is compatible with a
spiral ordering pattern.  While the mean field treatment of this
large-$N$ approach seems to 
miss the predominant spiral ordering of the
chain spins, the spiral ordering of the interstitial spins is in agreement with the
findings of this work.

Numerical studies of the spatially anisotropic kagome model
should be very helpful in establishing the range of spatial
anisotropy of exchange interaction where the non-coplanar
ordered state found in this work represents the ground state
of the system. We hope that our work will inspire further
investigations of this interesting problem.

\acknowledgments 

The authors thank J.-S.~Caux for sharing his numerical data on
the dynamical structure factor of the Heisenberg antiferromagnetic
chain, and Ashvin Vishwanath and Fa Wang for conversations.  This work
was supported in part by the National Science Foundation under
Grant No.\ PHY05-51164.  A.P.S. thanks the Swiss National Science 
Foundation for its financial support.  L.B. was supported by a 
David and Lucile Packard Foundation
Fellowship and the NSF through DMR04-57440.

\appendix

\section{RG calculations}
\label{app:rg-calculations}

In this Appendix we present a detailed derivation of flow equations
(\ref{eq:4}) and (\ref{eq:2}) using standard OPE relations for the
continuum fields.

\subsection{Derivation of Eqs.~(\ref{eq:4})}

The first order terms in the RG equations \re{eq:4} originate from the
rescaling of the time and space coordinates and the redefinition of the fields
[see Eqs.~\re{eq:5}]. As already discussed in the main text, the interaction
terms $V_1$ and $V_2$, Eqs.~\re{eq:6}, are local in space, which means that their scaling dimensions
have to be compared with $1$, the dimensionality of time $\tau$. 
Interstitial spins have scaling dimension $0$, while $\bm{M}$ ($\bm{N}$)
fields are of scaling dimension $1$ ($1/2$).
Consequently, the scaling dimension of $V_1$ is $1$, while that of $V_2$ is $3/2$, from which
follows that the RG equation for the coupling $\gamma_1$ does not contain a linear term, 
while that for $\gamma_2$ starts with 
$(1-3/2) \gamma_2 = - \gamma_2/2$, in agreement 
with the first line of Eqs.~\re{eq:4}.

The second and third lines of Eqs.~\re{eq:4} describe  the flow of the fluctuation-generated 
interchain couplings  
$V_{\mathrm{ch}}$, Eq.~\re{eq: interchain}, and
$V_{\mathrm{ch}}^{(1)}$, Eq.~\re{eq:1},
which operate  in two-dimensional space-time. As a result, the scaling dimensions
of the effective interchain interactions have to be compared with $2$. The first order
terms in the second line of Eqs.~\re{eq:4} are then a direct consequence
of the fact that the scaling dimension of  the
$\gamma_{\scriptscriptstyle\partial N}$-interaction is $3$, while that of the $\gamma_M$-interaction is $2$.
Similarly, positive linear terms in the last line of Eqs.~\re{eq:4} follow from the {\em strong relevance}
of the  interchain couplings $V_{\mathrm{ch}}$, Eq.~\re{eq: interchain}, which have scaling dimension $1$.

The second-order corrections to flow equations \re{eq:4}
are derived from contracting  terms in perturbation expansion~\re{eq: pert exp}.
In order to obtain the one-loop corrections in Eqs.~\re{eq:4}, we need to consider contributions 
to the second order term in Eq.~\re{eq: pert exp}  that either
yield a renormalization of the couplings $\gamma_i$ or generate new \emph{interchain}
interactions.
We begin by selecting from these contributions terms that contain a product of two intermediate
spins from the same site, $(2x+\frac{1}{2}, y + \frac{1}{2} )$, say. 
These \emph{local} contributions read 
\begin{eqnarray} \label{eq: terms for Vch}
&& \hspace{-0.2 cm}
\frac{1}{2} \mathrm{T}   \int d \tau_1 d \tau_2  \,
 s^a_{y+ \frac{1}{2} } ( \tau_1 )
s^b_{y+ \frac{1}{2} } ( \tau_2 )
\\
&& \times
\Bigg(
\gamma^2_1 
\left[ M^a_y ( \tau_1 ) + M^a_{y+1} (\tau_1) \right]
\left[ M^b_y ( \tau_2 ) + M^b_{y+1} (\tau_2) \right]
\nonumber\\
&& +
2 \gamma_1 \gamma_2 
\left[ M^a_y ( \tau_1 ) + M^a_{y+1} (\tau_1) \right]
 \partial_x
\left[ N^b_y ( \tau_2) + N^b_{y+1} ( \tau_2) \right]
\nonumber\\
&& +
\gamma^2_2   \partial_x
\left[ N^a_y ( \tau_1 ) + N^a_{y+1} ( \tau_1 ) \right]
\partial_x
\left[ N^b_y ( \tau_2) + N^b_{y+1} ( \tau_2) \right]
\Bigg) ,
\nonumber
\end{eqnarray}
where we have suppressed the $x$ coordinate for brevity.
Even though the interchain spins ${\bf  s}_{y+ \frac{1}{2} }$ have no dynamics
of their own at this level, they must be time-ordered as follows \cite{affleck1991}
\begin{eqnarray} \label{ap:spin comm}
&& \mathrm{T} s^a_{y+ \frac{1}{2} }( \tau_1 ) s^b_{y+ \frac{1}{2} } ( \tau_2 ) 
\nonumber\\
&& \quad =
\theta_{\tau_1-\tau_2} s^a (\tau_1 ) s^b (\tau_2 )
+ \theta_{\tau_2 -\tau_1 } s^b (\tau_2) s^a (\tau_1)  
\nonumber\\
&& \quad = 
\frac{\delta^{ab}}{4} + \frac{i}{2}
(\theta_t -  \theta_{-t}) 
\epsilon^{abc} s^c (\tau) ,
\end{eqnarray}
where $\theta_t$ is the step function,
$\tau = (\tau_1+\tau_2)/2$ is the center-of-mass time, and $t=\tau_1 - \tau_2$
is the relative time.

The off-diagonal term in Eq.~\re{ap:spin comm} 
 (which is proportional to $\epsilon^{abc}$)
is responsible for
the renormalization of the Kondo-like couplings $\gamma_1$ and $\gamma_2$.
As an example we consider the renormalization of $\gamma_1$, which
comes from the second line in Eq.~\re{eq: terms for Vch}.
Separating slow and fast degrees of freedom we can apply 
OPE~\re{eq: OPEs} (and a similar expression for the left currents)
to the product of two spin currents at nearby points
[i.e., $M^a_y(\tau_1) M^b_y(\tau_2) \to i \epsilon^{abd} M^d_y(\tau)/(2\pi t)$].
Combining this with \re{ap:spin comm} leads to
\begin{equation} \label{g_1 zwei}
- \frac{ \gamma_1^2 }{4 \pi u}
\int d \tau \,
 s^c_{y+ \frac{1}{2} } ( \tau )
  M^c_y  ( \tau )
  \int_{\alpha < | t | < b \alpha} \frac{d t}{|t|} .
\end{equation}
The integral with $| t | > b \alpha$ does not contribute to the renormalization.
The one-loop correction to the flow equation for $\gamma_1$ can now be read off \re{g_1 zwei} as
$\propto d\ell \gamma_1^2/\pi u$, which gives us the first
equation in \re{eq:4}.

The renormalization of $\gamma_2$ is computed in a similar way, one only needs to realize
that it comes from the third line in Eq.~\re{eq: terms for Vch}.
Fusing $M^a_y$   with $\partial_x N^b_y$  
via the OPE (see Ref.~\onlinecite{starykh2005} for more details)
\begin{eqnarray}
&&M^a_y(\tau_1) \partial_x N^b_y(\tau_2) = \lim_{x'\to x} M^a_y(x',\tau_1) \partial_x N^b_y(x,\tau_2)\nonumber\\
&&=\frac{-\delta^{ab} \varepsilon_y(x,\tau)}{2\pi (u t + a_0 \sigma_t)^2}  + 
\frac{i\epsilon^{abc} \partial_x N^c_y(x,\tau)}{2\pi (u t + a_0 \sigma_t)} ,
\label{ap:epsilon}
\end{eqnarray}
leads to the following one-loop correction
of the flow equation for $\gamma_2$: $\delta \gamma_2 \propto d\ell \gamma_1\gamma_2/\pi$.
It is useful to note that rescaling of space and time does not affect quadratic terms,
as each of them is explicitly proportional to the RG step $d\ell$, which comes
from the shell integration of relative coordinates.

The second-order corrections to the flow equations
for the interchain couplings $\gamma_M$ and $\gamma_{\partial N}$  follow from the diagonal 
term in Eq.~\re{ap:spin comm}  (which is proportional to $\delta^{ab}$).
For example, applying relation  \re{ap:spin comm} to the second line of Eq.~\re{eq: terms for Vch} produces
\begin{equation}
\frac{\gamma^2_1}{4 u} \int d\tau M^a_y(\tau) M^a_{y+1}(\tau)  \int_{\alpha < | t | < b \alpha} d t . \nonumber
\end{equation}
To generate from this the $\gamma_M$ term in Eq.~\re{eq:1} one needs to sum all local
contributions like the one above using $\sum_x ... = \int dx/(2a_0) ...$ , as
appropriate for the kagome geometry. As a result one finds that
$\gamma_M \sim \gamma_1^2 d\ell$.
Similarly, the other interchain coupling, $\gamma_{\partial N}$  in Eq.~\re{eq:1}, can be
derived starting from the last line in Eq.~\re{eq: terms for Vch}.

Finally, we turn to the relevant interchain interactions $\gamma_{\scriptscriptstyle N}$ and $\gamma_\varepsilon$ which
are generated by  fusing the marginal interaction $\gamma_M$ and the irrelevant interaction 
$\gamma_{\scriptscriptstyle\partial N}$. Details of this procedure have been described previously
in Refs.~\onlinecite{starykh2004,starykh2005}.
Here we would only  like 
to mention that the RG scheme that we have adopted here
(i.e., integrating the one-loop $x$ integrals over the {\em entire} space of  relative $x$ coordinates while restricting
the relative time integral to the shell, $\alpha < |t| < b \alpha$), which is different
from the RG scheme of Refs.~\onlinecite{starykh2004,starykh2005}, 
does not modify the outcome
of the calculation in Refs.~\onlinecite{starykh2004,starykh2005} in any significant way. 
Namely, upon fusing $M^a_y(x_1, \tau_1)$
with $\partial_x N^b_y(x_2,\tau_2)$ on chain $y$ (and, similarly, on chain $y\pm 1$),
one arrives at the following integral over the relative coordinate $x=x_1 - x_2$ and over
relative time  $t=\tau_1 - \tau_2$:
$ I \sim \int_{-\infty}^\infty dx  \int_{\alpha < | t | < b \alpha} dt (x^2 + t^2)^{-2} \sim (b-1) \alpha^{-2}$.
This explains the structure of the quadratic terms in the last line of  Eqs.~\re{eq:4}.

Additionally, we note that the third line in Eq.~\re{eq: terms for Vch}
in combination with the first term in Eq.~\re{ap:epsilon} results in a strongly relevant contribution
(scaling dimension $1/2$)
\begin{equation}
\frac{-6 \gamma_1 \gamma_2}{\pi a_0 u} \int d\tau \varepsilon_y (x,\tau) .
\label{ap:dimer}
\end{equation}
In the two-dimensional version of the spatially anisotropic kagome lattice
this term cancels out, since the summation over local bow-tie crossings (that is, over $x$) brings in
the factor $(-1)^x$ [which originates from the staggering factor in Eq.~\re{S in continuum}],
resulting in $\int d\tau dx (-1)^x \varepsilon_y (x,\tau) \to 0$. This is how 
symmetry \re{eq: third symmetry} manifests itself.

In contrast, for the kagome \emph{strip} of extension one in the 
$y$ direction, the staggering factor $(-1)^x$ does not appear, since
the bow-tie crossings are separated 
by \emph{two} lattice spacings.
As a consequence, the expression~\re{ap:dimer}
turns into $\int d\tau dx  \, \varepsilon_y (x,\tau)$, which implies spontaneous dimerization
of the kagome strip, an ordering pattern that does not reduce the translational symmetry.
This strongly relevant term was missed in previous analytical work \cite{azaria1998}.
Numerical studies, on the other hand, did find dimerized ground states \cite{white00}.

\subsection{Derivation of Eq.~(\ref{eq:2})}

\label{RG for dangling}

To derive flow equation (\ref{eq:2}) we need to
select from the second order term in Eq.~\re{eq: pert exp}
contributions that contain products of two different intermediate
spins with the same $y$ coordinate, $y + \frac{1}{2}$, say.
Among these, the most important contributions are those,
that involve products of uniform or staggered magnetizations from the same 
chain
\begin{eqnarray} \label{eq: exansion in V}
&&+
2 \times \frac{1}{2} \mathrm{T}  \hspace{-0.1cm}
\sum_{x_1, x_2} \int d \tau_1 d \tau_2 \;
s^a_{2x_1+\frac{1}{2} } (\tau_1 )s^b_{2x_2+\frac{1}{2} } (\tau_2 )
\nonumber\\
&&
\qquad
\times \Big[
\gamma_1^2 M^a  (2x_1, \tau_1) M^b  ( 2x_2, \tau_2 )
\\
&&
\qquad \quad \quad
+
\gamma^2_2 \partial_x N^a  (2 x_1, \tau_1 ) \partial_x N^b  ( 2 x_2, \tau_2 )
\Big] ,
\nonumber
\end{eqnarray}
where we have suppressed the $y$ coordinate for brevity.
The factor of $2$ in the first line arises because
there is an equal contribution from both the $y$ and
the $(y+1)$ chains. 
Since the spins $\bm{s}_{2x + \frac{1}{2} }$ at different sites
commute and time-ordering of the continuum fields $\bm{M}$ and $\bm{N}$
 is automatic, we can disregard the operator $\mathrm{T}$ in the
 above expression, provided we exclude the case $x_1 = x_2$, which was treated in
 the previous subsection.
 
By splitting the integrals of Eq.~\re{eq: exansion in V} into slow and fast degrees, 
we can use the OPEs, Eqs.~\re{eq: OPEs}, to fuse the product of 
two continuum fields at nearby points. 
In this way, we derive from Eq.~\re{eq: exansion in V}
the one-loop renormalization to the first term of the interaction $H_\triangle$, Eq.~\re{H dreieck}
\begin{subequations}
\begin{eqnarray}
 \sum_{x_1, x_2}
\int d \tau
s^a_{2x_1+\frac{1}{2} }  s^a_{2x_2+\frac{1}{2} }  
\left( \gamma^2_1 I^+_M + \gamma^2_1 I^-_M  + \gamma^2_2 I_N \right),
\end{eqnarray}
with the integrals
\begin{eqnarray}
\label{Ia}
I^{\pm}_M 
&=&
\int_{\alpha < |t | < b \alpha}  d t
\frac{  1 /( 8 \pi^2 )    }{ \left[u t \pm i ( x_1 - x_2 ) + a_0 \sigma_t \right]^2 } ,
\\
\label{Ib}
I_N
&=&
\int_{\alpha < |t | < b \alpha}  d t \,
\partial_{x_1} \partial_{x_2}
\frac{ C_N}{ \sqrt{u^2 t^2 + 4 (x_1 -x_2)^2 }} ,
\end{eqnarray}
\end{subequations}
where $C_N\approx (2\pi)^{-3/2}$ is the amplitude of
the $\langle N^a N^a \rangle$ correlator.
The integral over the infinitesimal interval $[ \alpha, b \alpha]$ in 
Eqs.~\re{Ia} and~\re{Ib} amounts to replacing t with $\alpha_\ell$.
The subsequent rescaling turns the cut-off $\alpha_\ell$ back into 
the microscopic cut-off $\alpha$,
while $x \to x_\ell = x e^{-\ell}$.
Simplifying $I_M = I_M^+ + I_M^{-}$ and explicitly taking derivatives in $I_N$
leads to Eqs.~\re{eq:2}.
It is worth noting that the two contributions $I_M$ and $I_N$ are of opposite signs,
resulting in a fast decay of interactions between
interchain spins~$\bm{s}$. 

\section{Chain ordering in the limit of a large spiral wave vector ($q \sim O[1]$) }  

\label{sec:large-q}

In this Appendix, we briefly discuss how the chain spins might order
under the perturbation of a spiral magnetic field with a large
wave vector $\bm{q}=(q,0)$, where $q \sim O[1]$. In this case the
transformation to a rotated frame as done in Section~\ref{sec:rotated
  frame} is not useful, as the generated interaction terms come with
couplings of the order of the bare exchange $J$, see Eqs.~\re{eq:H1} and
\re{eq:H2}.  Instead, we should remain in the original, non-rotated
basis, which leaves the dominant $O(J)$ interactions in their simplest
form.  Moreover, the rapid oscillation of the spiral field, which is
highly incommensurate with the ``natural'' wave vectors $0$ and $\pi$ 
of correlations of the Heisenberg chains, ensures that its effects rapidly
average out to leading order.  However, to $O(h^2)$ we may expect
non-oscillating interactions to be generated.  In principle these can be
obtained by a perturbative analysis expanding in powers of $H_S$
\re{spiral field}.

As usual, symmetry analysis is helpful in figuring out the type of terms
that can be expected from such a calculation.  There are two important
symmetries.  First, the spiral field term breaks both spin-rotational
and translational symmetries, but is invariant under a translation
$x\rightarrow x+1$ {\sl followed by} a simultaneous spin rotation by the
angle $q$ about the $z$ axis.  Spiral state \re{eq:s-spiral} is also
invariant under spatial inversion ($x \to - x$) followed by a change of
sign for the $y$ and $z$ components of the spins ($S^{y,z} \to
-S^{y,z}$).  Taking into account these constraints, the only possible
marginal or relevant terms which may be generated in a single chain are
\begin{equation}
  \label{eq:8}
  H_{h}^{(2)} = \sum_{xy} \int\! dx\, \left\{ d_z (J_{yR}^z-J_{yL}^z) + \delta g_z J_{yR}^z
    J_{yL}^z\right\} .
\end{equation}
We neglect here terms that are already present without the spiral field,
and those that couple different chains as these coupling
constants are necessarily smaller by at least one power of $J'$, since
the latter cannot be generated without some bare inter-chain
interactions.  

A naive current algebra calculation using the continuum approximation,
Eq.~(\ref{S in continuum}) for the spin operators, indeed produces
precisely these terms at $O(h^2)$.  However, the resulting explicit
expressions are not reliable, as they are dominated by short distances
of order $q^{-1}$, in which the continuum limit is inappropriate.  A
lattice scale calculation, similar in spirit to that in
Sec.~\ref{sec:numer-estim-inter} (yet more closely to that in
Ref.\cite{stoudenmire2007}), is required.  Unfortunately, it turns out
that to do so requires detailed information on the lattice scale
properties of certain {\sl four-spin} correlation functions of a
Heisenberg chain.  These data are not available to our knowledge.
Therefore, we must rely upon the symmetry considerations alone, assuming
no particular signs or magnitudes for $d_z$ and $\delta g_z$ above,
apart from the fact that both are expected to be of $O(h^2/J)$.

Fortunately, this does not result in significant ambiguity.  This is
largely because the nominally ``relevant''  DM correction $d_z$ has
trivial effects.   Similar to the magnetic field
in \re{eq: field shift} it can be removed by a simple shift. Unlike the
case in Sec.~\ref{subsec:rg}, however, the shift does not affect the
backscattering $H_{\text{bs}}$, which is written in terms of the field
$\varphi$, which is dual to $\theta$.  The only effect of the shift is
to change the ordering momentum, if any, of the $N^{x,y}$
components.  It does not determine the nature of the ordering
instability.  

The anisotropy term $\delta g_z$ {\sl is} important, as it tips the
balance of  competition between different components of $\bm{N}$
fields in the interchain Hamiltonian $H''$, Eq.~\re{eq:H-2prime}. With
the help of an OPE-based calculation similar to the one that led to
Eq.~\re{eq:final-rg} we find 
\begin{equation}
  \label{eq:9}
  \frac{d}{d\ell} \ln
  \frac{\gamma_{N^z}}{\gamma_{N^x}} = \frac{\delta g_z}{4\pi u} .    
\end{equation}
This tells us that the type of $\bm{N}$-order is determined by the {\sl
  sign} of the generated $\delta g_z$.  Positive $\delta g_z$ favors
$N^z$ components, leading to the non-coplanar ordering pattern found in
Section~\ref{sec:order-chain}, in the small-$q$ limit.
Negative $\delta g_z$, on the other hand, would prefer $N^{x,y}$
components, without breaking the symmetry between them. Such a state is
clearly {\sl co-planar}, and different from the one found in
Section~\ref{subsec:rg}.  Note that the DM term $d_z$ will affect the
ordering wave vector of this state but not the non-coplanar one. Which of
these two situations is obtained cannot be discriminated by our analysis,
since the sign of $\delta g_z$ is not determined.  However, it is rather
natural to expect that as $q$ is reduced from $O(1)$ values, one should
observe behavior consistent with the small $q$ analysis.  This would
suggest that $\delta g_z>0$ for a non-vanishing range of $q$ greater
than zero, and indeed it is possible that this is the case for all $q$.

\end{document}